\documentclass[twocolumn]{aastex61}

\newcommand\aastex{AAS\TeX}

\usepackage{amsmath}
\accepted{July 1, 2019}

\submitjournal{AAS Journals}

\shorttitle{\aastex\ Four Low-mass {\it Kepler} Eclipsing Binaries}
\shortauthors{Han et al.}

\begin{document}

\title{Magnetic Inflation and Stellar Mass IV. Four Low-mass {\it Kepler} Eclipsing Binaries Consistent with Non-magnetic Stellar Evolutionary Models}

\correspondingauthor{Eunkyu Han}
\email{eunkyuh@bu.edu}

\author[0000-0001-9797-0019]{Eunkyu Han}
\affil{Department of Astronomy \& Institute for Astrophysical Research, Boston University, 725 Commonwealth Avenue, Boston, MA 02215, USA}

\author[0000-0002-0638-8822]{Philip S. Muirhead}
\affil{Department of Astronomy \& Institute for Astrophysical Research, Boston University, 725 Commonwealth Avenue, Boston, MA 02215, USA}

\author[0000-0002-9486-818X]{Jonathan J. Swift}
\affiliation{The Thacher School, 5025 Thacher Rd. Ojai, CA 93023, USA}

\begin{abstract}
Low-mass eclipsing binaries show systematically larger radii than model predictions for their mass, metallicity and age. 
Prominent explanations for the inflation involve enhanced magnetic fields generated by rapid rotation of the star that inhibit convection and/or suppress flux from the star via starspots. However, derived masses and radii for individual eclipsing binary systems often disagree in the literature. In this paper, we continue to investigate low-mass eclipsing binaries (EBs) observed by NASA's {\it Kepler} spacecraft, deriving stellar masses and radii using high-quality space-based light curves and radial velocities from high-resolution infrared spectroscopy.  We report masses and radii for three {\it Kepler} EBs, two of which agree with previously published masses and radii (KIC 11922782 and KIC 9821078).  For the third EB (KIC 7605600), we report new masses and show the secondary component is likely fully convective ($M_2 = 0.17 \pm 0.01 M_{\sun}$ and $R_2 = 0.199^{+0.001}_{-0.002} R_{\sun}$). Combined with KIC 10935310 from Han et al. (2017), we find that the masses and radii for four low-mass {\it Kepler} EBs are consistent with modern stellar evolutionary models for M dwarf stars and do not require inhibited convection by magnetic fields to account for the stellar radii.

\end{abstract}




\section{Introduction}\label{intro}
Double-lined eclipsing binary stars (SB2 EBs) offer a powerful method to empirically determine stellar masses and radii through photometric and spectroscopic observations. Photometric data allow the determination of the radius ratio, the sum of the radii (in units of the semi-major axis), and the surface brightness ratio, which is often converted into a temperature ratio using atmospheric models.  With high signal-to-noise eclipse photometry, the orbital eccentricity and argument of periastron can be determined directly from the light curve.  Spectroscopic radial velocity measurements of both stars allow the determination of the physical scale of the system through the measurement of the semi-major axis and individual component masses. \\ \indent Empirically determined masses and radii are critical to both stellar astrophysics and exoplanet studies. The measurements are essential to test the detailed astrophysics of stellar evolutionary models. SB2 EBs with at least one low-mass main-sequence ($M_{\star} \lesssim 0.7 M_{\sun}$) star are useful for testing the treatment of convection and degeneracy in stellar evolutionary models \citep[e.g.][]{Feiden2013}.
Moreover, the properties of M dwarf exoplanet host stars need to be characterized accurately to understand their exoplanet populations.\\
\indent Although EBs offer a direct way to empirically determine mass-radius relationship of M dwarf stars, only a few dozen low-mass EBs are known \citep[][]{Torres2010, Feiden2012}, and the measurements show large scatter around model predictions. The measured M dwarf radii differ by 5 to 10\% on average for their mass and age. Some M dwarf stars seem to have hyper-inflated radii that is offset by 100 to 200\% (e.g. NSVS 02502726 \citep{Lee2013}, T-Lyr0-08070 \citep[][]{Cakirli2013b}, and CSSJ074118.8+311434 \citep{Lee2017}). Theoretical efforts have been undertaken to fix discrepancies between observations and model predictions.  For instance, PARSEC (the Padova and Trieste Stellar Evolutionary Code) is a revised version of the Padova evolutionary model \citep[][]{PARSEC}, which incorporated updated input physics (e.g. stellar opacities, equation of state) and microscopic diffusion in low-mass stars in order to fix the mass-radius discrepancy. \\
\indent There have been different scenarios proposed to explain the discrepancies between the empirical measurements and the model predictions. A prominent theory for the inflated radii involves enhanced magnetic fields from rapid rotation of the star, where strong magnetic fields on the order of several kilo-Gauss are sustained in the stellar atmosphere, which inhibit convection \citep[][]{Chabrier2007}. This effect depends largely on the mass of the star, with higher-mass stars are more affected than lower-mass stars. Moreover, enhanced magnetic fields produce surface spots, hindering the radiative loss of heat at the surface. When the stellar surface is covered by more spots given the same effective temperature, the effective temperature is effectively reduced, resulting in a larger radius for the same mass and luminosity. Indeed, in a previous paper in this series, \citet{Kesseli2018} presented evidence that fully convective, rapidly rotating single M dwarf stars do in fact appear 10-to-15\% larger than evolutionary models predict for their absolute $K$-band magnitudes, supporting the starspot hypothesis.
\indent Another possible explanation for inflated radii involves challenges associated with acquiring high quality data and eclipse fitting. Currently, astronomers have discovered dozens of low-mass EBs, but there are outstanding questions about the role of the analysis in determining the parameters and the quality of data. For example, in the studies of KIC 10935310, an M dwarf EB with {\it Kepler} photometry, \citet[][]{Cakirli2013a} found the secondary component was inflated and the primary was not, whereas \citet[][]{Iglesias2017} found the primary was inflated and the secondary was not. However, in a previous paper in this series, \cite{Han2017} measured the mass and the radius of each component that differ significantly from the previous two measurements. The results are broadly consistent with modern stellar evolutionary models for main-sequence low-mass stars and do not require inhibited convection by magnetic fields to account for the stellar radii. The differences in measured parameters were attributed to the differences in the quality of the radial velocity data. The first two groups used moderate-resolution optical spectra, where \cite{Han2017} used high-resolution near-infrared spectra. \cite{Kraus2017} and \cite{Gillen2017} independently studied AD 3814, a low-mass EB in Praesepe, and measured different parameters for the secondary component. \cite{Gillen2017} found a radius that is consistent with model predictions where \cite{Kraus2017} found an inflated radius.\\
\indent To measure stellar parameters accurately, we need high fidelity photometric and spectroscopic data. NASA`s {\it Kepler} Mission measured near-continuous light curves for hundreds of thousands of stars over four years with the aim of discovering Earth-sized exoplanets transiting Sun-like stars \citep[e.g.][]{Kepler_Citation}. {Hundreds of eclipsing binaries have since been found in the {\it Kepler} light curves \citep[e.g.][]{Prsa2011}, though few have spectroscopic measurements. For low-mass EBs spe}cifically, high-resolution near-infrared spectra are powerful in determining the masses of individual components in EBs, and for measuring stellar radii in physical units. There are two major advantages of spectroscopy in the near-infrared. Measurements in the near-infrared are less sensitive to stellar activity (e.g. starspots). Starspots on a rotating photosphere can introduce radial velocity variations \citep[e.g.,][]{Andersen2015}. In the near-infrared, the spot-induced radial velocity signal is significantly reduced due to the lower contrast between spots on the photosphere at longer wavelengths \citep[][]{Reiners2010}. M dwarf stars are also brighter in infrared than at optical wavelengths, providing a higher signal to noise ratio. \\
\indent In this work, we investigated three {\it Kepler}  SB2 EB systems (KIC 11922782, KIC 9821078, and KIC 7605600) and measured the masses and radii of their component stars using a consistent approach, using {\it Kepler} data and high-resolution near-infrared spectra from IGRINS, iSHELL, and NIRSPEC. KIC 11922782 and KIC 9821078 have previous measurements by \citet[][]{Helminiak2017} and \citet[][]{Devor2008b}, and we find our measurements consistent with the literature. We also announce a new measurement of a low-mass SB2 EB system, KIC 7605600, which contains a fully-convective (M $\leq$ 0.33M$_\odot$) secondary M dwarf component. KIC 7605600 was first discovered and identified as an EB by 
\citet[][]{KeplerDR2} and was classified as M+M detached eclipsing binary by \citet[][]{Shan2015}. In their study of measuring the binarity of M dwarfs using the {\it Kepler} eclipsing binary data, \citet[][]{Shan2015} searched a set of M dwarf targets that were identified by \citet[][]{DR2013} and came up with 12 M+M eclipsing binaries, one of which was KIC 7605600. No previous work was done on characterizing the component stars. \\
\indent As we show in the following sections, we determined the masses and the radii of individual components of all three {\it Kepler} EBs. In Section \ref{Data} we describe the data used in our determinations. In Section \ref{model} we describe our modeling procedure and results. In Section \ref{Discussion} we discuss the implications for the masses and radii in comparison to the stellar evolutionary models.

%%%%%%%%%%%%%%%%%%%%%%%%%%%%%%%%%%%%%%%%%%%%%%%%%%%%%%%%%%%%%%%%%%%%%%%%%%%%%%%%%%%%%%%%%%%%%%
\section{Data and Observations}\label{Data}

\subsection{{\it Kepler} Light Curve}\label{lc}
\begin{figure*}
\centering
\includegraphics[width=7.5in]{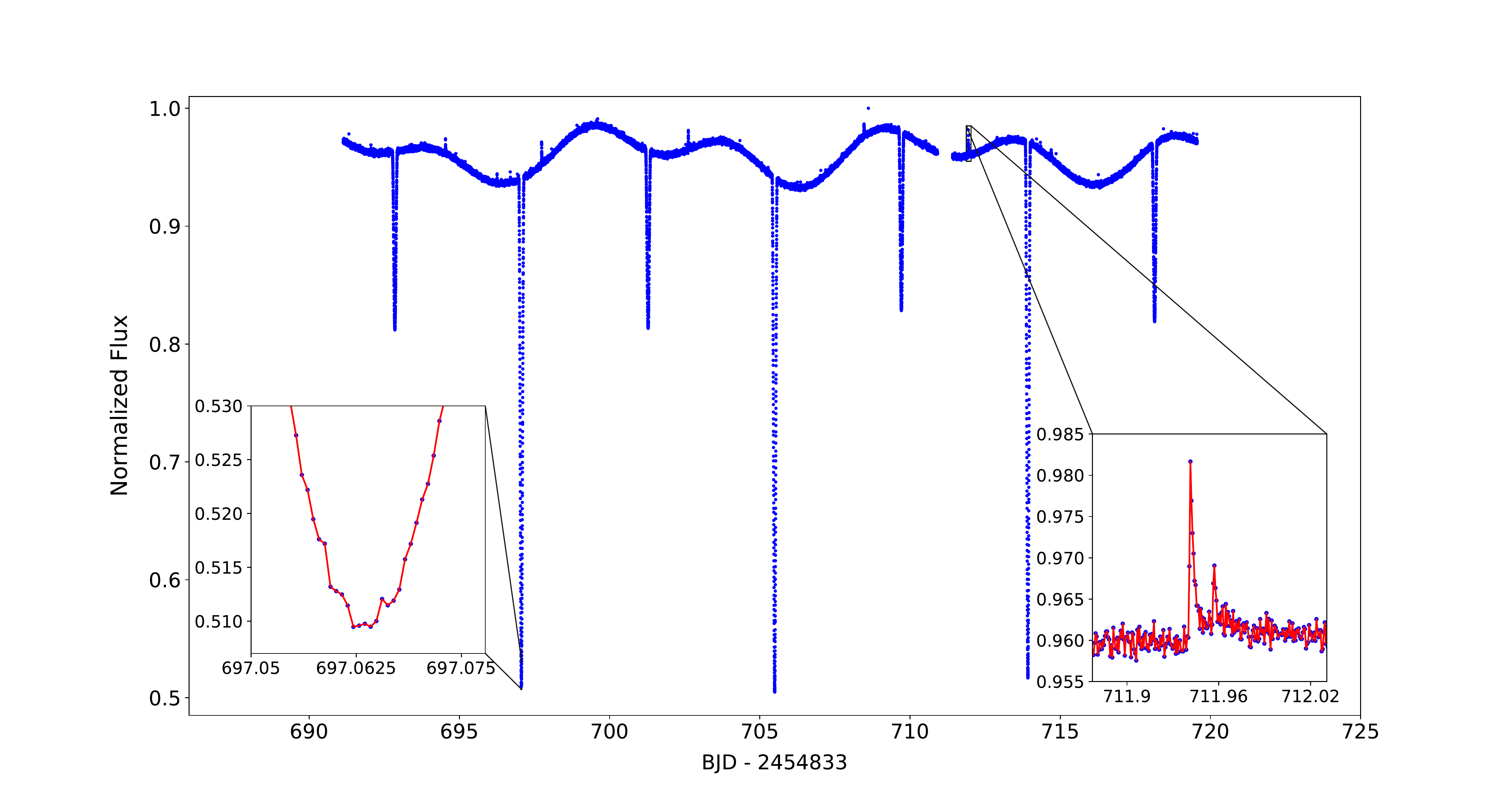}
\caption{Example of the {\it Kepler} short-cadence data showing eclipses of KIC 9821078 from quarter 7. The panel on the lower left corner contains a closer look at the primary eclipse. The other panel on the lower right corner contains a closer look at an abrupt increases in flux, likely caused by flares on the photosphere of either the primary or secondary component. The out-of-eclipse flux modulation is also shown, which we attribute to rotating star spots on the photosphere of either component star. The modulation period is consistent with star spots and spin-orbit synchronous rotation of either the primary or secondary component star, or both. The 1-min exposure times of {\it Kepler} short-cadence data provide ample coverage across each individual eclipse event.}
\label{kic9821078_sample_lc}
\end{figure*}

\begin{figure*}
\centering
\includegraphics[width=7.5in]{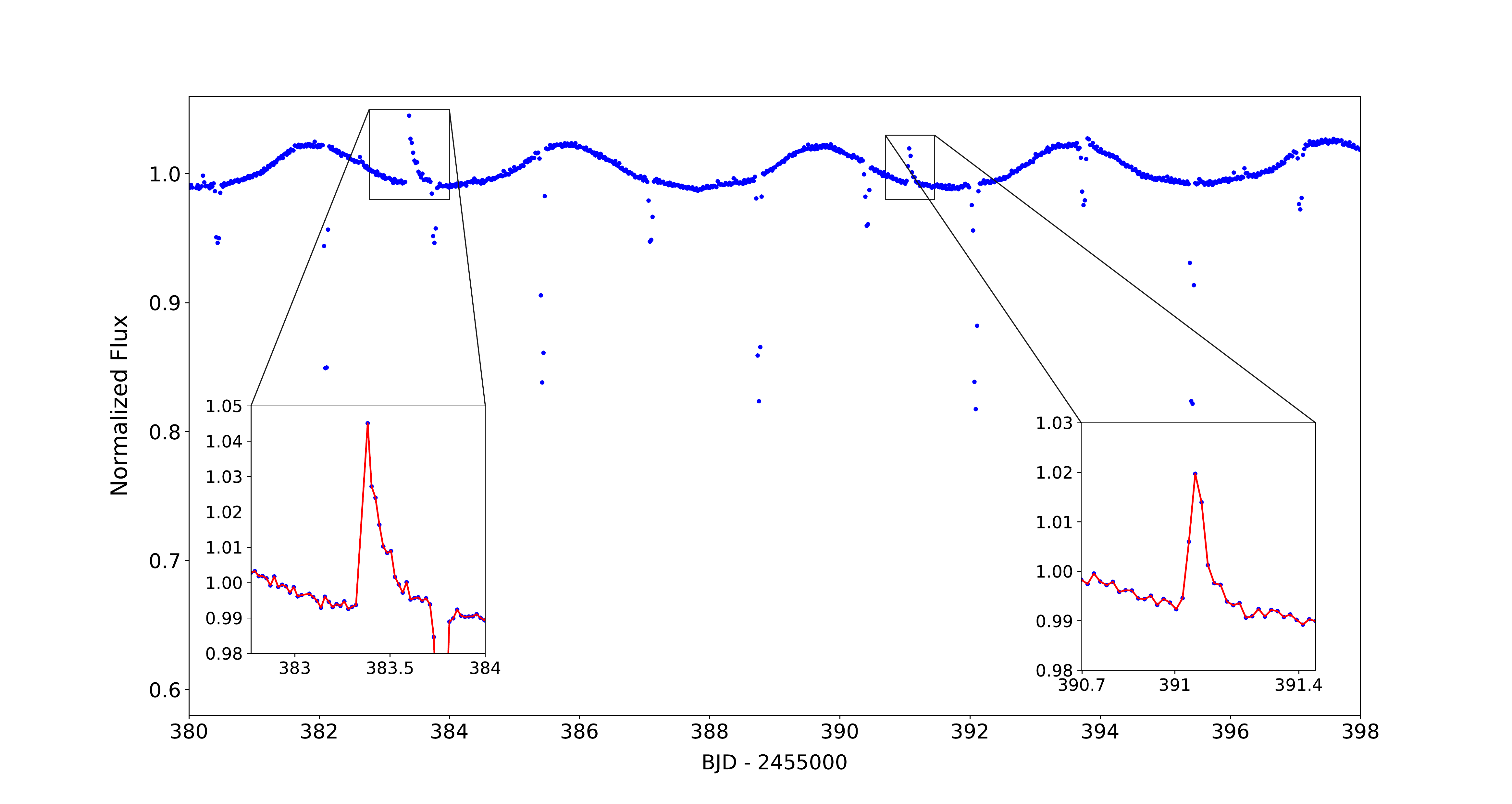}
\caption{Example of the {\it Kepler} long-cadence data showing eclipses of KIC 7605600 from quarter 6. Two small panels on the bottom of the figure contain a closer look at the abrupt increases in flux, caused by flares on the photosphere of either the primary or the secondary component. The out-of-eclipse flux modulation is also shown, which is caused by rotating star spots on the photosphere. The modulation period is consistent with star spots and spin-orbit synchronous rotation of either the primary or secondary component star.}
\label{kic7605600_sample_lc}
\end{figure*}

For all three systems, we obtained {\it Kepler} light curve data from the Mikulski Archive for Space Telescopes (MAST). \footnote{\url{https://dx.doi.org/10.17909/T9059R}}  Long-cadence data recorded at regular intervals and with exposure times of 1765.5 seconds are available for all quarters of the primary {\it Kepler} mission except for KIC 7605600, where only even numbered quarters from 2 through 16 are available. Short-cadence data recorded at regular intervals and with exposure times of 58.89 seconds are available for specific quarters for KIC 11922782 and KIC 9821078, with no short-cadence data available for KIC 7605600. We used the PDCSAP\_FLUX data, which is corrected for effects from instrumental and spacecraft variation \citep[][]{Stumpe2012, Smith2012}. The summary of all available {\it Kepler} data for the EB systems are shown in Table \ref{kepler_summary}. On inspection, all three EB systems' light curves show out-of-eclipse modulation that is nearly synchronous with the system orbital period. We attribute the modulation to starspots on the component stars combined with synchronous stellar rotation.\\
\indent Figure \ref{kic9821078_sample_lc} shows an example of the {\it Kepler} short-cadence data showing eclipses of KIC 9821078 from quarter 7. The 1-min exposure times of {\it Kepler} short-cadence data provide ample coverage across each individual eclipse event. Figure \ref{kic7605600_sample_lc} shows the same but of the {\it Kepler} long-cadence data of KIC 7605600 from quarter 6. The long-cadence data also captures the out-of-eclipse flux modulation, which is consistent with star spots and spin-orbit synchronous rotation of either the primary or secondary component star. \\
\indent We carried out an analysis on the short- and the long-cadence data independently and found that the measurements agree with each other. However, for all our analysis and reported parameters, we use the measurements from the long-cadence data for a consistent approach, since KIC 7605600 does not have short-cadence data. \\

\begin{table*}
\begin{center}
\caption{Available quarters of {\it Kepler} data for all three EB systems.}
\begin{tabular}{@{}l c c c c c c}
\hline
Target & RA & DEC & Gaia source ID (DR2) & Distance (pc) & Long Cadence & Short Cadence\\
\hline
KIC 11922782 & 	19$^h$44$^m$01$^s_.$770 & +50$^{\circ}$13$'$57 375$''$ & 2135341298718849536 & 235.864 & 1-17	&	2, 3 \\
KIC 9821078 & 19$^h$07$^m$16$^s_.$618 & +46$^{\circ}$39$'$53 150$''$ & 2130535195954075648 & 243.666 & 1-17	&	7, 8, 9, 10 \\
KIC 7605600 & 19$^h$24$^m$36$^s_.$150 & +43$^{\circ}$17$'$07 136$''$ & 2126014141581543424 & 159.746 &  2, 4, 6, 8, 10, 12, 14, 16	&	- \\
\hline

\hline
\label{kepler_summary}
\end{tabular}
\end{center}
\end{table*}

\subsection{SB2 Radial Velocity Data}
\subsubsection{IGRINS Observations}
\begin{figure}
\begin{center}
\includegraphics[width=\linewidth]{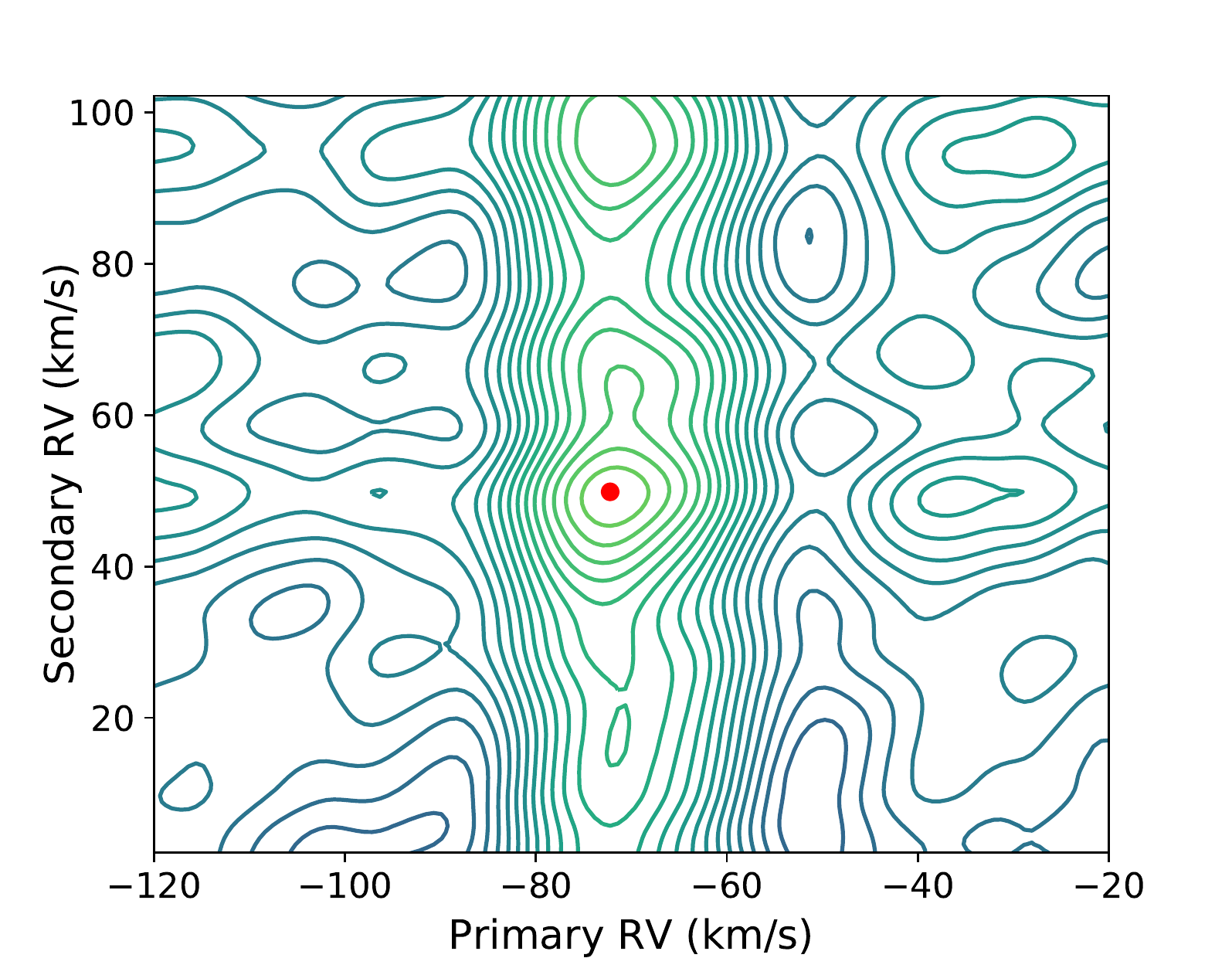}
\caption{A sample contour plot of the two-dimensional cross-correlation function using one of KIC 7605600's IGRINS spectrum. The lighter the color, the higher the two-dimensional cross-correlation function. The red dot indicates the location of the maximum value of the two-dimensional cross-correlation function. The uncertainty was estimated by calculating the standard deviation of the radial velocities across the orders and divided by the square root of number of orders used. }
\label{sample_todcor}
\end{center}
\end{figure}

We observed all three EB systems using the the Immersion GRating INfrared Spectrometer \citep[IGRINS,][]{Yuk2010} on the 4.3-meter Discovery Channel Telescope (DCT) in September, October, and November of 2017 and 2018. IGRINS is a cross-dispersed, high-resolution near-infrared spectrograph. The wavelength coverage is from 1.45 to 2.5 $\mu m$ and with a spectral resolution of R = $\lambda/\Delta\lambda$ = 45,000. IGRINS allows simultaneous observations of both H- and K-band in a single exposure \citep[][]{Yuk2010, Park2014, Mace2016}. For each science target, the exposure times were calculated to achieve a signal-to-noise ratio of $\sim$75 or higher per wavelength bin.  We also observed A0V standard stars within 0.2 airmass of the science targets for telluric corrections. For all our targets, we performed ABBA nodding. To reduce the spectra, we used publicly available reduction pipeline for the IGRINS \citep[][]{plp}. \\
\indent The pipeline performs dark subtraction, flat-fielding, AB subtraction to remove the OH airglow emission lines, and extracts the spectrum. We further processed the pipeline extracted 1-D spectra to correct any residuals from the telluric correction, which could affect our RV measurements. For this task, we used \texttt{xtellcor\_general}, a generalized version of SpeX's telluric correction software, \texttt{xtellcor}, designed to remove telluric lines from  near-infrared spectra \citep[][]{xtellcor}. The software takes an observed spectrum of an A0V star and a target spectrum, constructs the telluric spectrum from a model spectrum of Vega, calculates the relative shift between the two input spectra, and applies the shift to the constructed telluric spectrum. \texttt{xtellcor\_general} divides the constructed telluric spectrum from the target spectrum. \\
\indent IGRINS H- and K-band data contain 28 and 25 orders, respectively. However, we only used the H-band data for two reasons. The sky background in the K-band data are higher, reducing the signal-to-noise of the spectra. Furthermore, the current pipeline is known to show 2-3 $\,km\,s^{-1}$ of scatter due to a problem with distortion correction in the K-band. Of the 28 orders in the H-band spectra we selected 6$^{th}$, 7$^{th}$, and 11$^{th}$ through the 21$^{st}$ due to their high signal-to-noise ratios. These orders gave us a wavelength coverage of 1.49 $\mu m$ to 1.73 $\mu m$. For the radial velocity standards, we used BT-Settl model spectra \citep[][]{Allard2012} with different temperatures. The specifics of the model spectra are listed in Table \ref{BTSettl_detail} and can be obtained from the PHOENIX website.\footnote{https://phoenix.ens-lyon.fr/Grids/BT-Settl} The BT-Settl models were matched to have the same resolution as the IGRINS spectra but were not corrected for the rotational broadening. To measure the radial velocities, we first interpolated the spectra onto a logarithmic wavelength scale to make the sampling uniform in velocity space. We used the Two-dimensional CORrelation technique \citep[TODCOR,][]{TODCOR} and calculated the radial velocities of each component. We calculated the radial velocity for each order separately. We adopted the mean of the radial velocities returned for each order as the measured radial velocity, and adopted the uncertainty by calculating the standard deviation of the radial velocities across the orders and dividing by the square root of number of orders used. The detailed procedure can be found in \citet[][]{Han2017}. The top panel in Figure \ref{sample_spectra} shows a sample IGRINS H-band telluric-corrected spectrum of KIC 7605600 (in blue) and two BT-Settl spectra (in red and green). Figure \ref{sample_todcor} shows a sample contour plot of the two-dimensional cross-correlation function using one of KIC 7605600's IGRINS spectrum. The lighter the color, the higher the two-dimensional cross-correlation function. The red dot indicates the location of the maximum value of the two-dimensional cross-correlation function.

\begin{table*}
\begin{center}
\caption{Details of the BT-Settl models used.}
\begin{tabular}{@{}l c c c}
\hline
Target & T$_{eff}$  & Metallicity & logg\\
\hline
KIC 11922782 & 	5800K, 6000K & 0.0 & 5.0 \\
KIC 9821078 &  3300K, 4000K & 0.0 & 5.0 \\
KIC 7605600 & 3000K, 3800K & 0.0 & 5.0  \\
\hline
\label{BTSettl_detail}
\end{tabular}
\end{center}
\end{table*}

\subsubsection{NIRSPEC Observations}
We observed KIC 7605600 with NIRSPEC on the W. M. Keck II Telescope \citep[][]{McLean1998} on the UT nights of 2014 July 6, 13, and 19. NIRSPEC is a cross-dispersed near-infrared spectrograph that gives a spectral resolution of R = $\lambda/\Delta\lambda$ = 25,000. KIC 7605600 was observed in the K-band using the NIRSPEC-7 filter, which covers the wavelength of 1.839 to 2.630 $\mu m$, with an ABBA nodding pattern. We observed A0V standard stars on each night that are within 0.2 airmasses of KIC 7605600 for the purpose of telluric corrections. \\
\indent We reduced the data using \texttt{REDSPEC}, a publicly available IDL based reduction pipeline for NIRSPEC \citep[][]{REDSPEC}. \texttt{REDSPEC} subtracts dark exposures, divides by flat-field exposures, rectifies each frame, performs the AB subtraction, and extracts 1-D spectrum. We further processed the pipeline extracted 1-D spectrum with a custom script to correct the wavelength solution as for some orders, the arc lamp did not give enough prominent lines to precisely determine the wavelength solution. The custom script uses the ATRAN model of telluric lines \citep[][]{ATRAN} to compare the observed telluric absorption lines in the A0V spectra. We calculated shifting and stretching parameters of the wavelength solution by minimizing the $\chi^2$ of wavelength corrected A0V and the ATRAN model. Any corrections in the wavelength solution were then applied to both the observed A0V spectra and KIC 7605600's. After the wavelength corrections, we used \texttt{xtellcor\_general} and performed the same procedure as we did for the IGRINS data to remove telluric lines. We also found an additional NIRSPEC observation of KIC 9821078 from the nights in July and August of 2006 and July of 2007 on the Keck Observatory Archive (KOA)\footnote{https://koa.ipac.caltech.edu/cgi-bin/KOA/nph-KOAlogin} and have included them in our analysis. \\
\indent We performed the same method as we did with IGRINS data to calculate the radial velocity. The middle panel in Figure \ref{sample_spectra} shows a sample NIRSPEC K-band telluric-corrected spectrum of KIC 7605600 (in blue) and two BT-Settl spectra that are matched to have the same spectral resolution as that of NIRSPEC (in red and green). \\

\subsection{iSHELL Observations}
\begin{figure*}
\begin{center}
\includegraphics[width=0.85\linewidth]{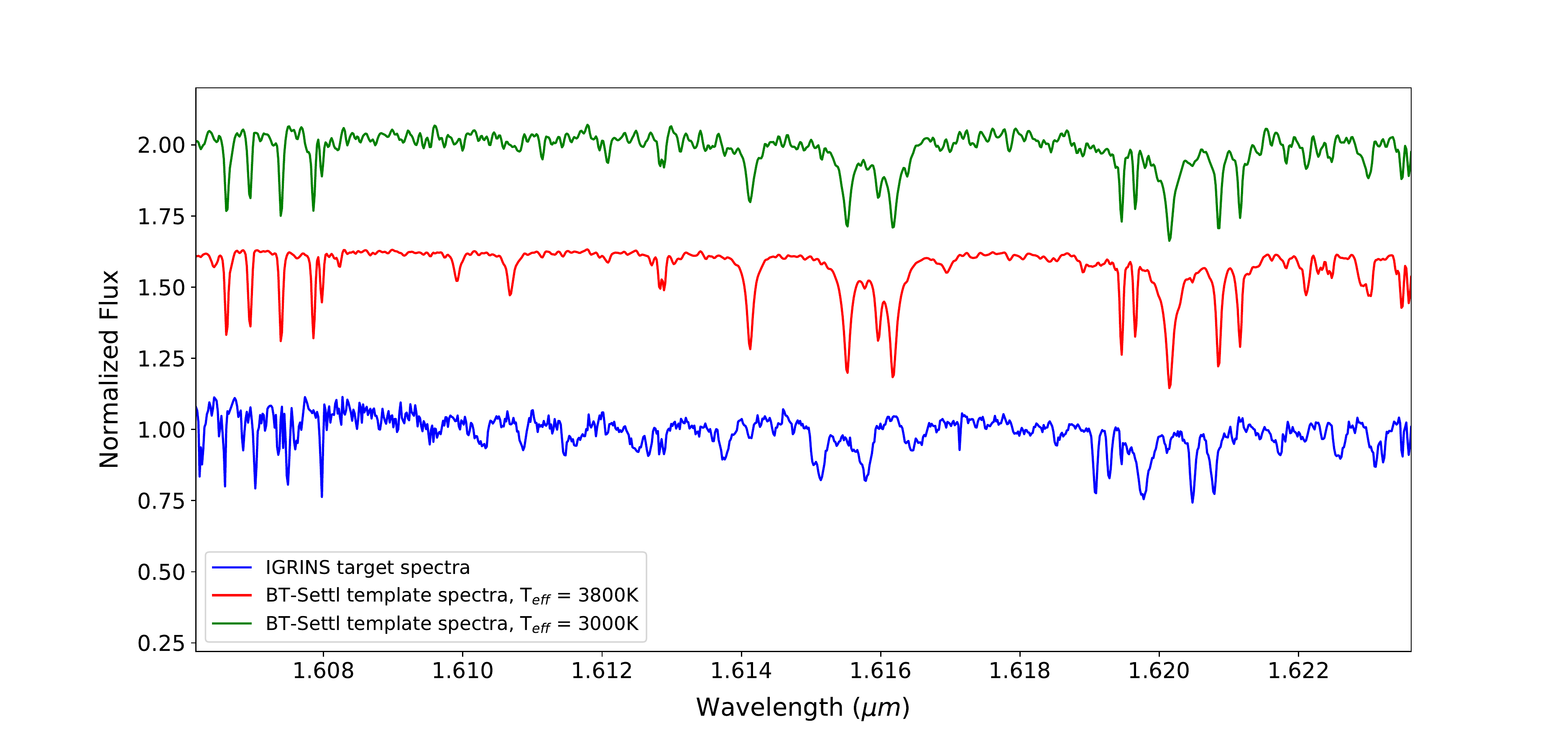}
\includegraphics[width=0.85\linewidth]{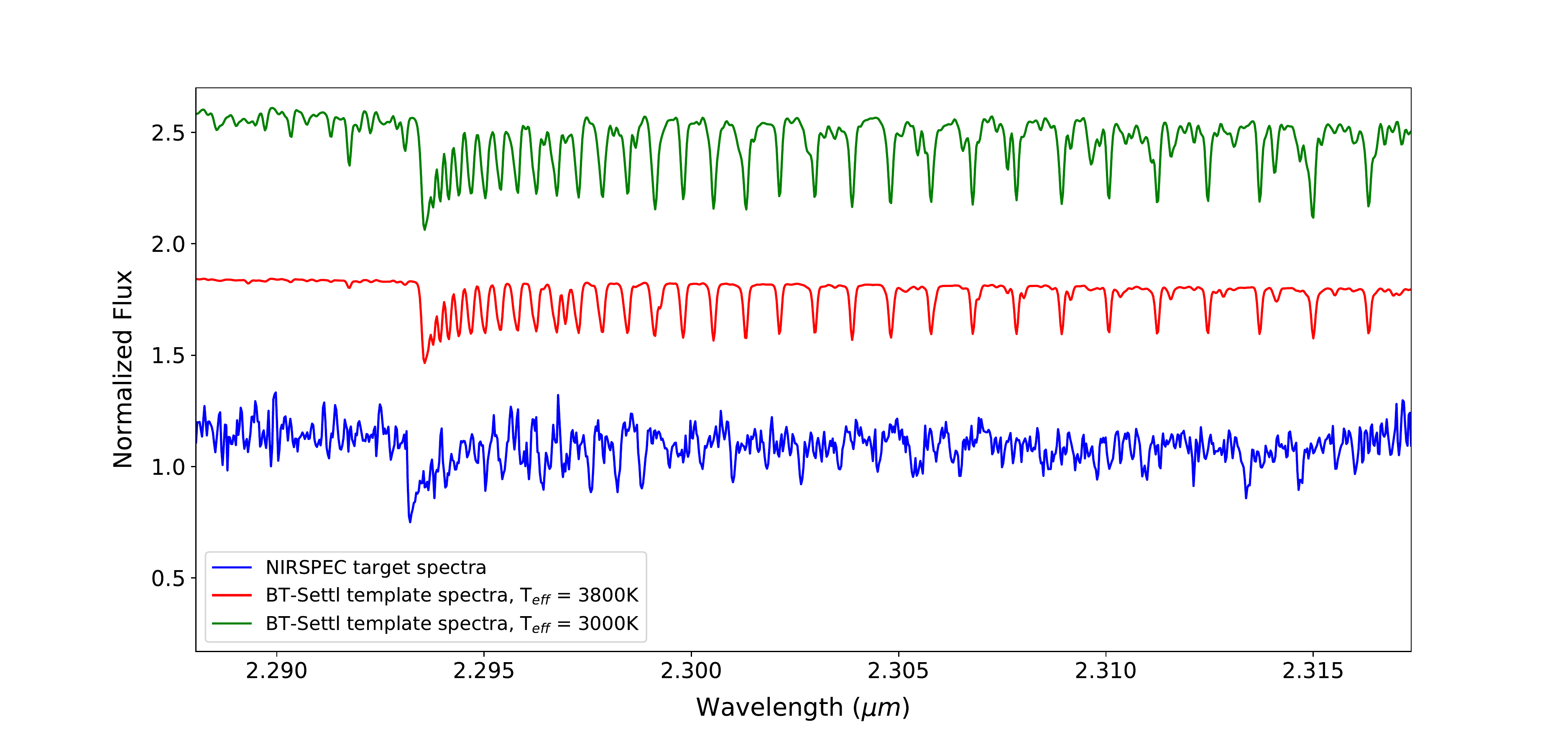}
\includegraphics[width=0.85\linewidth]{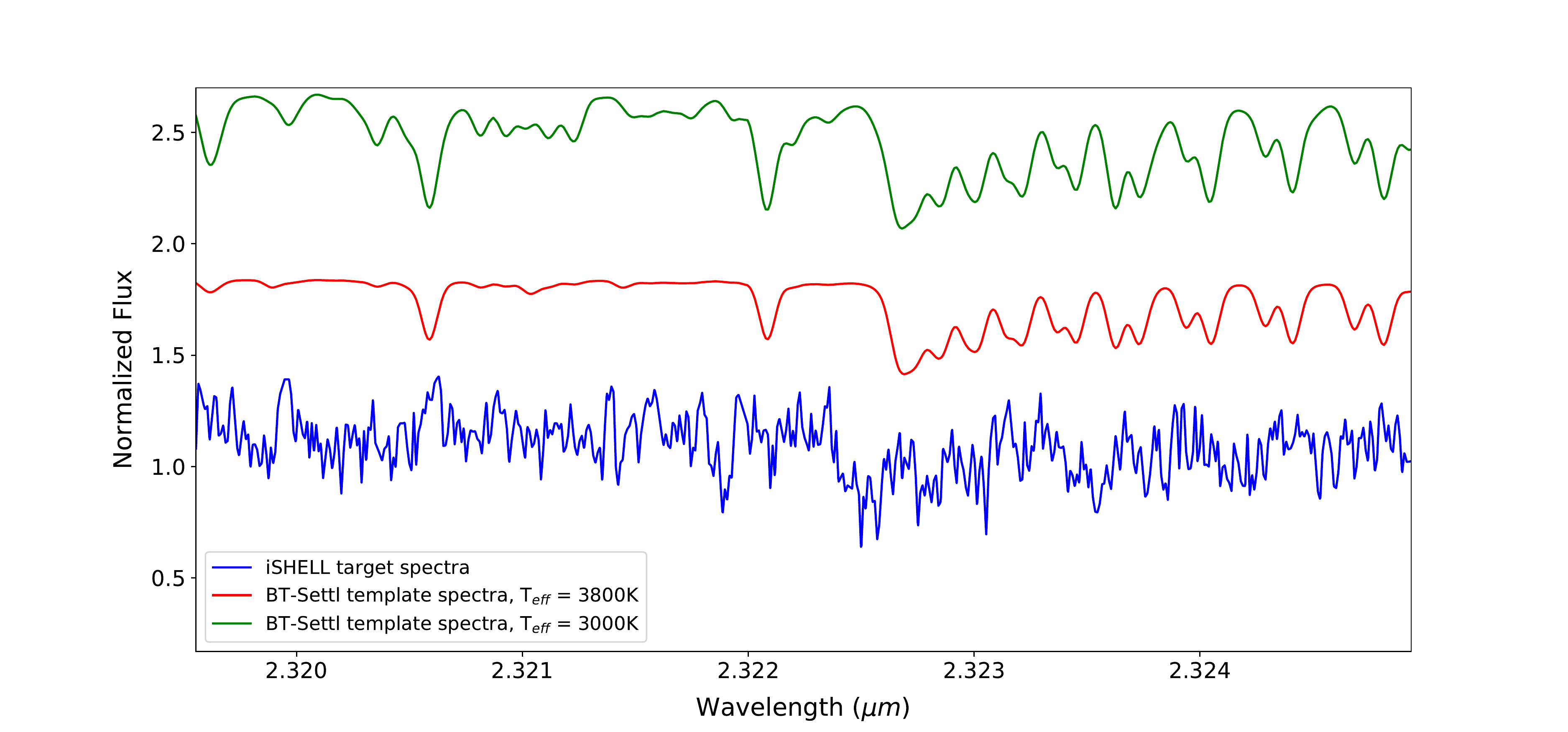}

\caption{Example spectra of KIC 7605600 from IGRINS (top), NIRSPEC (middle), and iSHELL (bottom). BT-Settl spectra that are used as the radial velocity templates are plotted for comparison (red and green). Each plot shows a single order from the respective instrument and BT-Settl spectra matching the spectral resolution of the respective instrument. 
Due to the RV offsets, the target spectra are shifted with respect to the BT-Settl spectra.}
\label{sample_spectra}
\end{center}
\end{figure*}

We observed KIC 9821078, KIC 9641031, and KIC 7605600 using iSHELL on NASA's InfraRed Telescope Facility (IRTF) on the nights in September of 2017 and June and August of 2018. iSHELL is a cross-dispersed near-infrared spectrograph that covers $\sim1.1 \mu m - 5.3 \mu m$, with two options of slit width that give resolving powers of  $R=\lambda/\Delta\lambda = 35,000$ and  $R=\lambda/\Delta\lambda = 75,000$. We used the K2 filter, which covers from 2.09 $\mu m$  to 2.38 $\mu m$. We aimed for radial velocity precision of 3\% or better and signal-to-noise of $\sim$75 or higher per wavelength bin. We used the resolution of 35,000 over 75,000 because the calculated exposure times for 75,000 would cause shifts of spectral lines from the motion of the stars during observations. For each science observation, we took  calibration observations which include dome flats and arc lamp as well as the A0V standards as required for iSHELL. \\
\indent We reduced iSHELL data using the iSHELL version of \texttt{Spextool} \citep[][]{spextool} and telluric corrected using \texttt{xtellcor}. We removed any obvious outliers (hot or otherwise bad pixels) by masking and interpolating across them. iSHELL's K2-band data have 29 orders and we only used the orders from 4 through 8, and 15.  To calculate the radial velocities, we performed the same method as we did with IGRINS data. The bottom panel in Figure \ref{sample_spectra} shows a sample iSHELL K-band telluric-corrected spectrum of KIC 7605600 (in blue) and two BT-Settl spectra that are matched to have the same spectral resolution as that of iSHELL (in red and green). \\

\subsection{Radial Velocity data from literature}
For KIC 11922782 and KIC 9821078, which have previously been studied by \citet[][]{Helminiak2017} and \citet[][]{Devor2008}, respectively, we also took the published radial velocity measurements and combined with our measurements. For the measurements from \citet[][]{Devor2008}, the authors report the uncertainty in the range of 0.5 and 1.2 $\,km\,s^{-1}$ and we used the larger uncertainty in order to be conservative. Furthermore, the NIRSPEC data we found on KOA were identical to the ones used by \citet[][]{Devor2008} for their analysis, and our independently determined radial velocity measurements were consistent. 

%%%%%%%%%%%%%%%%%%%%%%%%%%%%%%%%%%%%%%%%%%%%%%%%%%%%%%%%%%%%%%%%%%%%%%%%%%%%%%%%%%%%%%%%%%%%%%

\section{Analysis and Results}\label{model}

\subsection{Light Curve Model and Fit}\label{LCmodel}
To analyze {\it Kepler} data, we followed the approach of \citet[][]{Han2017}. We first modeled the out-of-eclipse modulations in long-cadence and short-cadence data, separately, using \texttt{george}, a Gaussian processes module written in Python \citep[][]{george}. The best-fit out-of-eclipse model obtained from \texttt{george} is then divided out of {\it Kepler} data. After detrending, we normalized the flux by dividing by the median value of the out-of-eclipse portion. We also rejected any outliers in the out-of-eclipse that are 2$\sigma$ above or below the median value. To model the detrended light curves, we used \texttt{eb}, a publicly available eclipsing binary modeling code written for detached eclipsing binaries \citep[][]{Irwin2011}. The model takes 37 free parameters of which the 16 parameters of interest are described in Table \ref{modelparm}. \\
\indent We modeled the long- and short-cadence data separately. We first explored the long-cadence data by searching for the best-fit model through employing the Levenberg-Marquardt technique and performing $\chi^2$ minimization, using Python's external package, \texttt{mpfit} \citep[][]{Markwardt2009}. We further refined the fit and determined the uncertainties for each individual parameters by employing the Markov chain Monte Carlo (MCMC) algorithm, using Python's external MCMC package, \texttt{emcee} \citep[][]{Foreman-Mackey2013}. The best-fit parameters from \texttt{mpfit} were used to set the starting parameters in the MCMC chains. We employed 500 walkers, each with 8000 steps, and assumed uniform priors on all parameters. We explored all parameters listed in Table \ref{modelparm} and treated them as free parameters. A special note for $L3$ is that unlike the case in KIC 10935310, where high contrast imaging was available to directly determine the contribution of the third light to the system total flux, we lack high-contrast imaging data for these targets. We visually inspected UKIDSS images and did not see indications of third body for both KIC 9821078 and KIC 11922782. For KIC 7605600, we were not able to rule out the third body and let $L3$ be explored by the MCMC chains. The extracted stellar parameters from varying $L3$ were consistent with those from fixed $L3$, except for the total flux level in the out-of-eclipse. We report the parameters from the MCMC with fixed $L3$.\\
\indent As shown in Table \ref{modelparm}, \texttt{eb} uses square-root limb-darkening law. We converted the square-root limb-darkening coefficients to $q_1$ and $q_2$, as developed by \citet[][]{Kipping2013}, stepped in the $q$s, and converted back to the square-root limb darkening coefficients for the model computations. Kipping $q_1$ and $q_2$ parameterization forces all possible combinations of $q_1$ and $q_2$ to be physical, as long as both values are between 0 and 1. As done in \cite{Han2017}, for $e\cos{\omega}$ and $e\sin{\omega}$, we stepped in $\sqrt{e} \cos{\omega}$ and $\sqrt{e} \sin{\omega}$, suggested by \citep[][]{Eastman2013}. Imposing uniform priors for $e\cos{\omega}$ and $e\sin{\omega}$ biases towards high values of eccentricity, as noted in \cite{Ford2006}.\\ 
\indent Once the MCMC algorithm finished exploring all possible parameter space for the long-cadence data, we further explored the model parameters using the short-cadence data. The set of model parameters that resulted the maximum likelihood value from the long-cadence was used as the starting parameters of the MCMC chains for the short-cadence data. \\
\indent When fitting, we smoothed the \texttt{eb} light curve model to account for the {\it Kepler} long- and short-cadence integration time. Furthermore, we excluded the majority of the out-of-eclipse for two reasons. They were the dominant noise source in the $\chi^2$  calculation and the flattened out-of-eclipse fluxes had no information on the physical parameters component stars except for the total flux.\\
\indent When the MCMC chains converged, we visualy inspected the chains, removed the first 3000 steps (the ``burn-in''), and took the most probable parameters from a single step in the chains where the likelihood was the maximum. We do not report the median of the posteriors but choose to report parameters from the single step with the highest likelihood to give a more accurate approximation of the posterior distributions. For the parameters with symmetric posterior distributions, we took the standard deviations of the MCMC chains as the uncertainties. For the parameters with asymmetric posterior distributions, we took the difference between the values of the maximum likelihood and the 34.1$^{th}$ percentile around the maximum likelihood and reported as asymmetric uncertainties.\\
\indent For all analysis, we focused on the long-cadence data to maintain the consistency in our measurements, since there was no short-cadence data for KIC 7605600. However, for the other two systems with the short-cadence data, we cross-checked the measurements from long- and short-cadence data to ensure they are consistent. 

\begin{table*}
\begin{center}
\caption{Modeling Parameters}
\begin{tabular}{@{}l c}
\hline
Parameter & Description \\
\hline
$J$ & Central surface brightness ratio \\
$(R_1 + R_2)/a$ & Fractional sum of the radii over the semi-major axis\\
$R_2/R_1$ & Radii ratio\\
$\cos{i}$ & Cosine of orbital inclination \\
$P$ (days) & Orbital period in days \\
$T_0$ (BJD) & Primary mid-eclipse time \\
$e \cos{\omega}$ & Orbital eccentricity $\times$ cosine of argument of periastron  \\
$e \sin{\omega}$ & Orbital eccentricity $\times$ sine of argument of periastron\\
$L3$ & Third light contribution\\
$\gamma$ (km s$^{-1}$) & Center of mass velocity of the system \\
$q$ & Mass ratio ($M_2$/$M_1$) \\
K$_{tot}$/c & Sum of the radial velocity semi-amplitude in units of c\\
LDLIN1 & Linear limb-darkening coefficient for the primary\\
LDNON1 & Square root limb-darkening coefficient for the primary\\
LDLIN2 & Linear root limb-darkening coefficient for the secondary\\
LDNON2 & Square root limb-darkening coefficient for the secondary\\
\hline
\label{modelparm}
\end{tabular}
\end{center}
\end{table*}

\subsection{Radial Velocity Model and Fit}
\indent Tables \ref{11922782_rv}, \ref{9821078_rv}, \ref{7605600_rv} show the measured radial velocities of KIC 11922782, KIC 9821078, and KIC 7605600, respectively. For all three systems, we measured radial velocities of both primary and secondary components using IGRINS and NIRSPEC spectra, except for one epoch of KIC 11922782. We only measured the primary radial velocity as shown in Table \ref{11922782_rv}, whose epoch was near 0.5 in orbital phase. However, using iSHELL spectra, we were not able to measure radial velocities of secondary components in most cases due to having lower signal-to-noise ratio that range between 25 to 35.
To measure the masses of each component, we fit the photometric and the spectroscopic data individually because the number of data points in the {\it Kepler} data far outweigh the radial velocity data, which would result in poor fit in the radial velocity data when fit globally. Instead of employing the MCMC algorithm to extract the individual masses, we linearized the radial velocity equation as a function of the radial velocity semi-amplitudes, $K_1$, $K_2$, and the systematic velocity, $\gamma$, and used an analytic fitter to calculate the best-fit parameters. The detailed derivation of the analytic fitter is in Appendix \ref{RVanalytic}. \\
\indent We attribute the possibility of analytic radial velocity fit to the high-fidelity {\it Kepler} data. Among the parameters that affect the radial velocity model, we determined the orbital period ($P$), the epoch of the primary mid-eclipse ($T_0$), $e\cos{\omega}$,and $e\sin{\omega}$ with high precision. For the fore-mentioned parameters, we took the most probable values from the MCMC chains of the long-cadence {\it Kepler} data and fit for $K_1$, $K_2$, and $\gamma$. \\

%%%%%%%%%%%%%%%%%%%%%%%%%%%%%%%%%%%%%%%%%%%%%%%%%%%%%%%%%%%%%%%%%%%%%%%%%%%%%%%%%%%%%%%%%%%%%%

\subsection{Results}
\begin{figure*}
\centering
\includegraphics[width=7.0in]{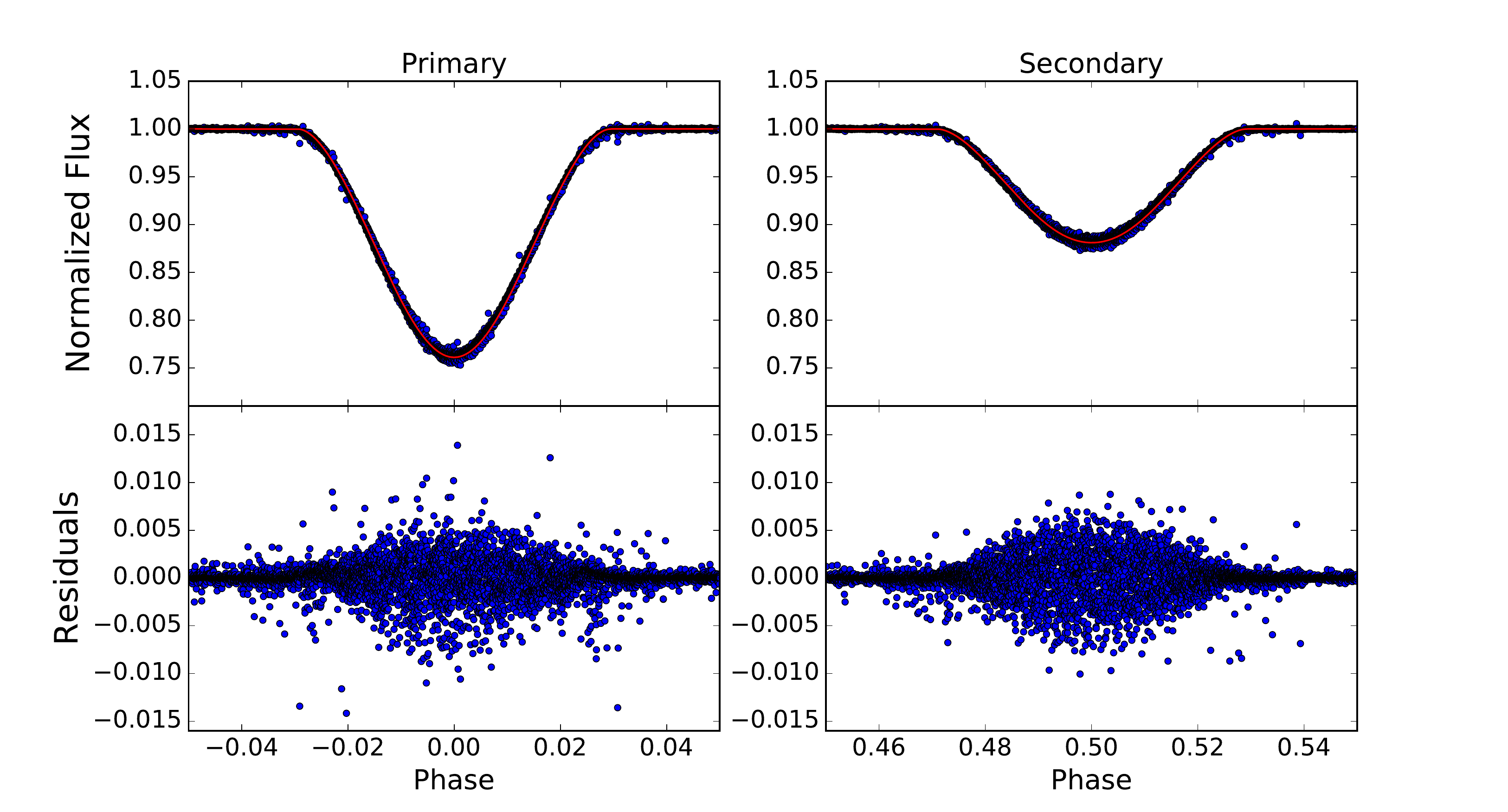}
\caption{Zoom-in of the phase-folded primary and secondary eclipses of KIC 11922782. The top panels show detrended and phase folded {\it Kepler} data with their best fit and the bottom panels show the residuals. We ascribe the scatter in the residuals to spot crossing events.}
\label{11922782_model_fit}
\end{figure*}

\begin{figure*}
\centering
\includegraphics[width=0.85\linewidth]{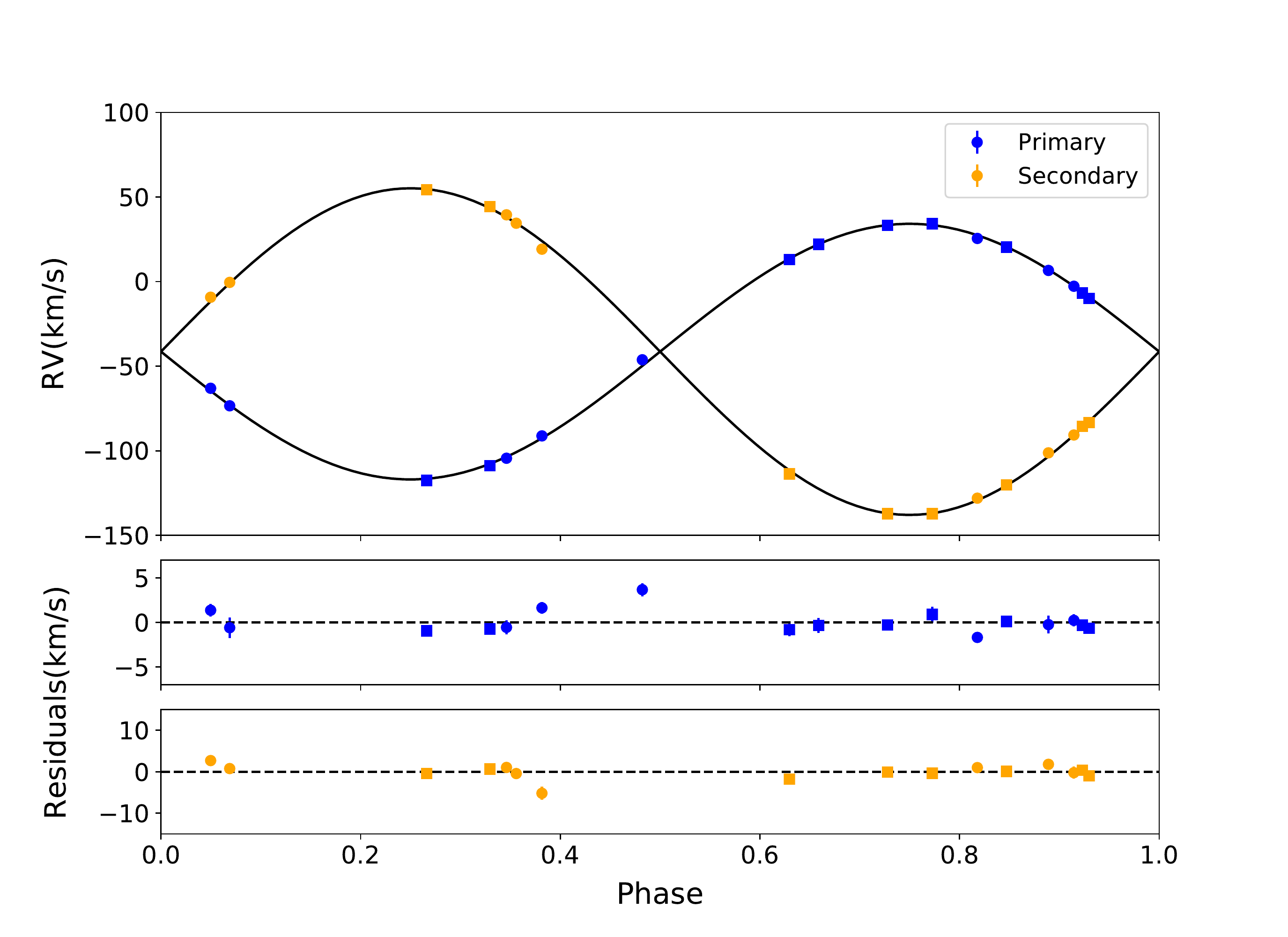}
\caption{Best-fit to the radial velocity data of KIC 11922782. In blue and in orange are the radial velocity data of the primary and the secondary, respectively. The circles denote the radial velocity measurements from this work and the squares are those from \citep[][]{Helminiak2017}. The solid black line is the analytically calculated best-fit model for all data. The bottom two panels show the residuals for each component with their corresponding colors. The calculated radial velocity semi-amplitudes are $K_1 = 75.6 \pm 0.1 \,km\,s^{-1}$ for the primary and $K_2 = 96.6 \pm 0.1 \,km\,s^{-1}$ for the secondary.}
\label{11922782_rvfit}
\end{figure*}

\begin{figure*}
\centering
\includegraphics[width=0.75\linewidth]{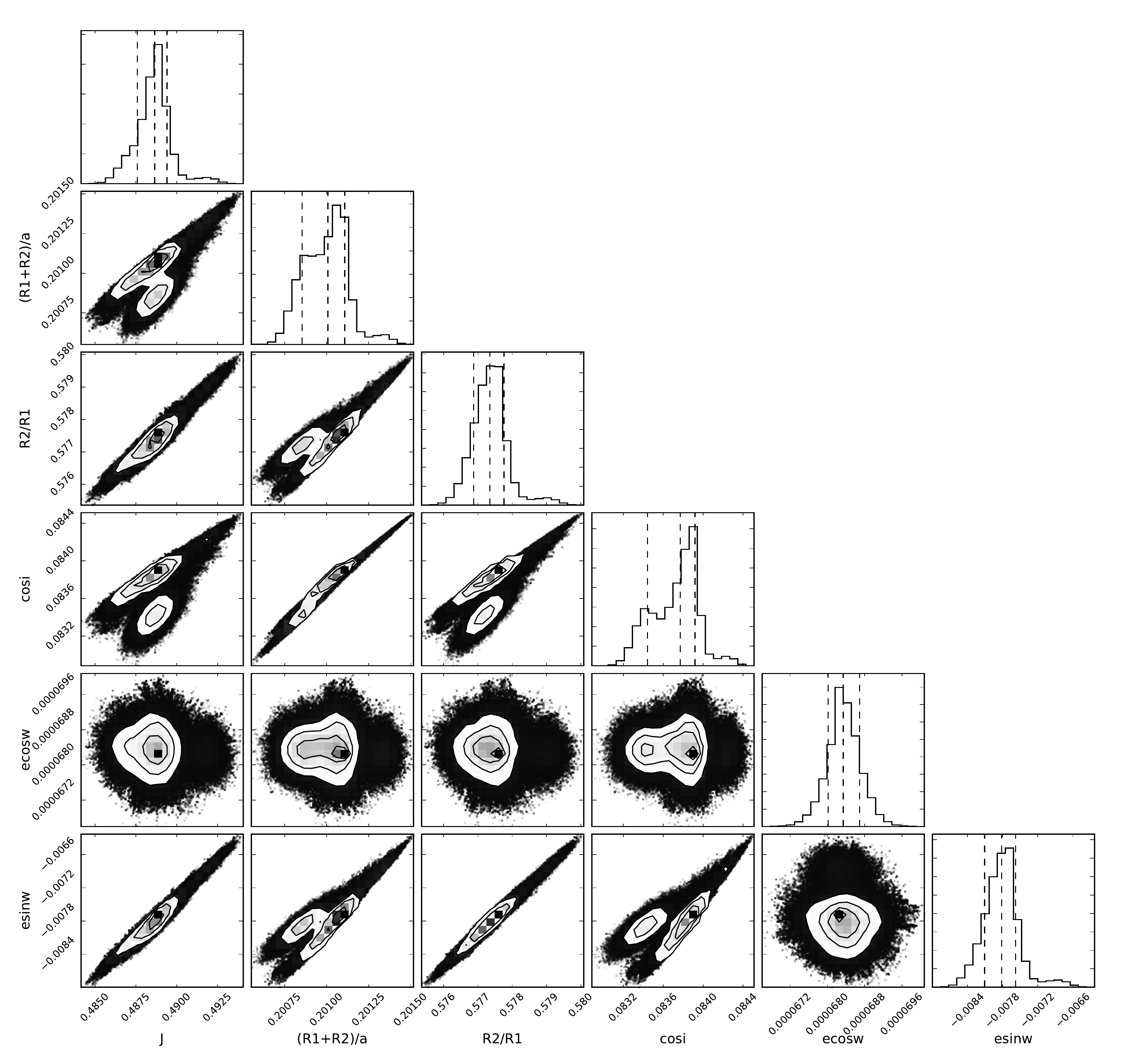}
\caption{Triangle plot of KIC 11922782 from the light curve fit. The histogram and the contour plots show density of MCMC iterations. The dashed lines in the histogram mark 16$^{th}$, 50$^{th}$, and 84$^{th}$ percentiles of the samples in the marginalized distributions. 
See Table \ref{modelparm} for descriptions of the fitted parameters.}
\label{11922672_LC_corner}
\end{figure*}

\begin{figure*}
\centering
\includegraphics[width=7.0in]{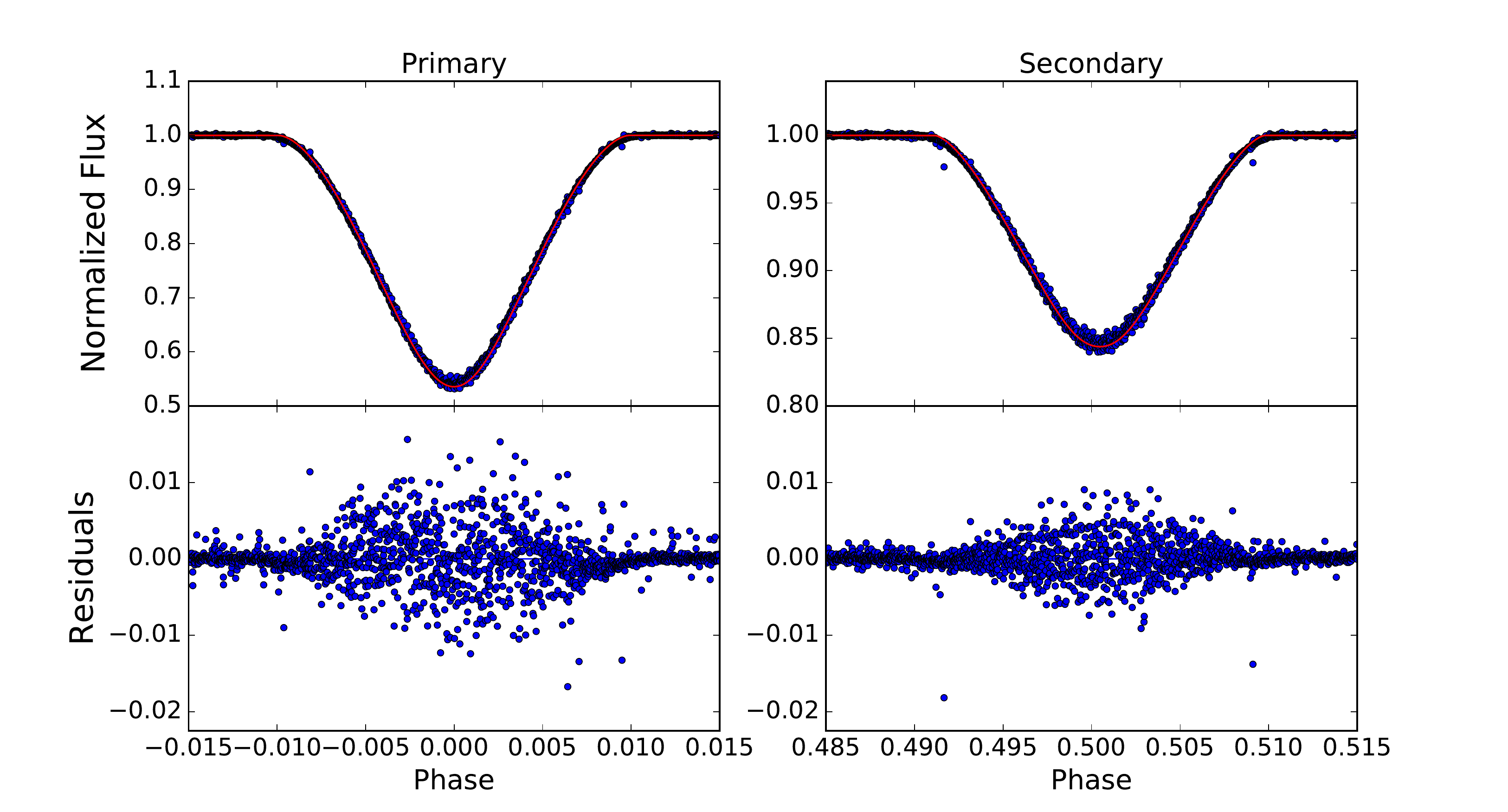}
\caption{Zoom-in of the phase-folded primary and secondary eclipses of KIC 9821078. The top panels show detrended and phase folded {\it Kepler} data with their best fit and the bottom panels show the residuals. We ascribe the scatter in the residuals to spot crossing events.}
\label{9821078_model_fit}
\end{figure*}

\begin{figure*}
\centering
\includegraphics[width=0.85\linewidth]{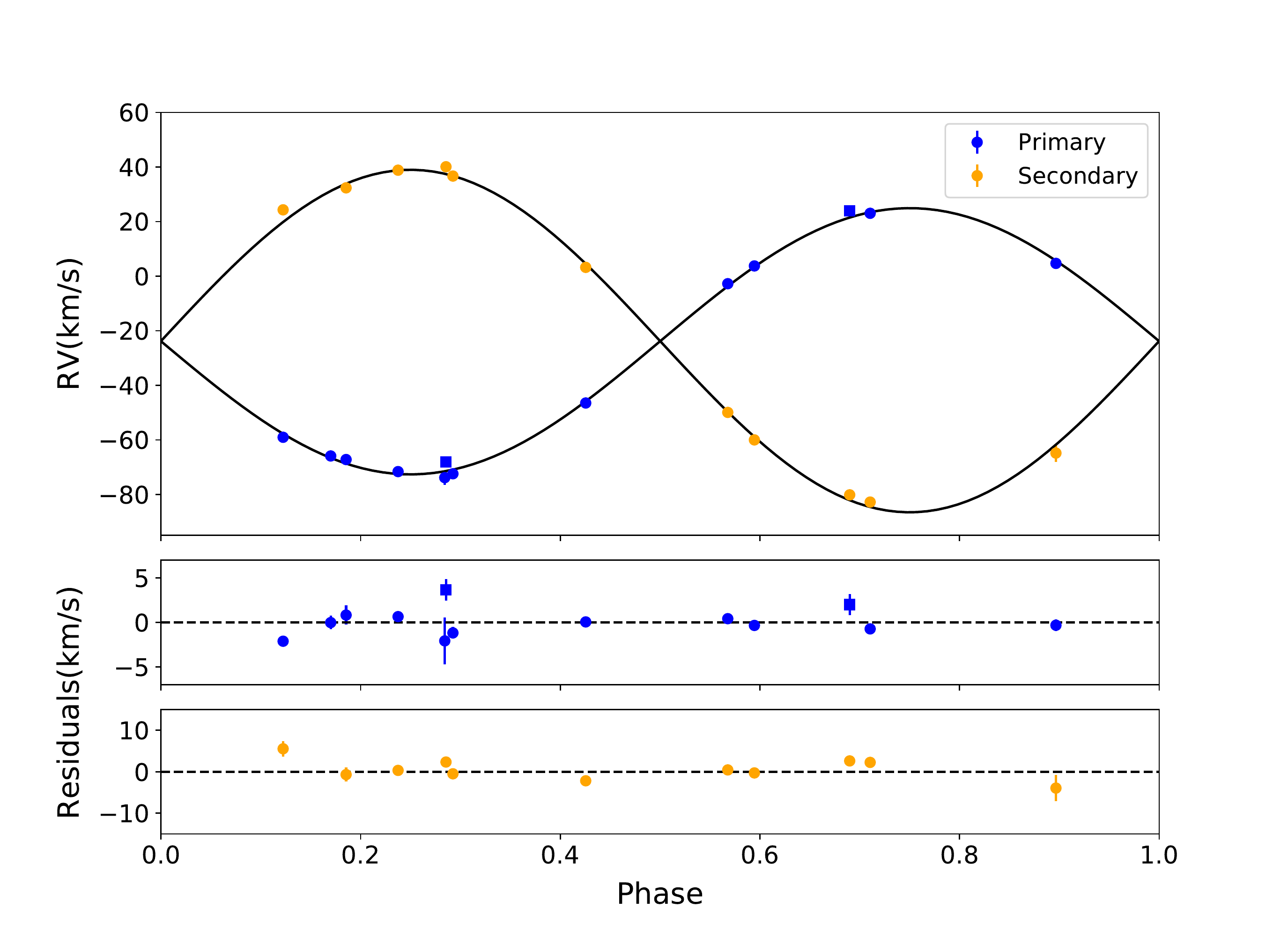}
\caption{Best-fit to the radial velocity data of KIC 9821078. In blue and in orange are the radial velocity data of the primary and the secondary, respectively. The circles denote the radial velocity measurements from this work and the squares are those from \citep[][]{Devor2008}. The solid black line is the analytically calculated best-fit model for all data. The bottom two panels show the residuals for each component with their corresponding colors. The calculated radial velocity semi-amplitudes are $K_1 = 48.8 \pm 0.1\,km\,s^{-1}$ for the primary and $K_2 = 62.8 \pm 0.1\,km\,s^{-1}$ for the secondary using all available data.} 
\label{9821078_rvfit}
\end{figure*}

\begin{figure*}
\centering
\includegraphics[width=0.75\linewidth]{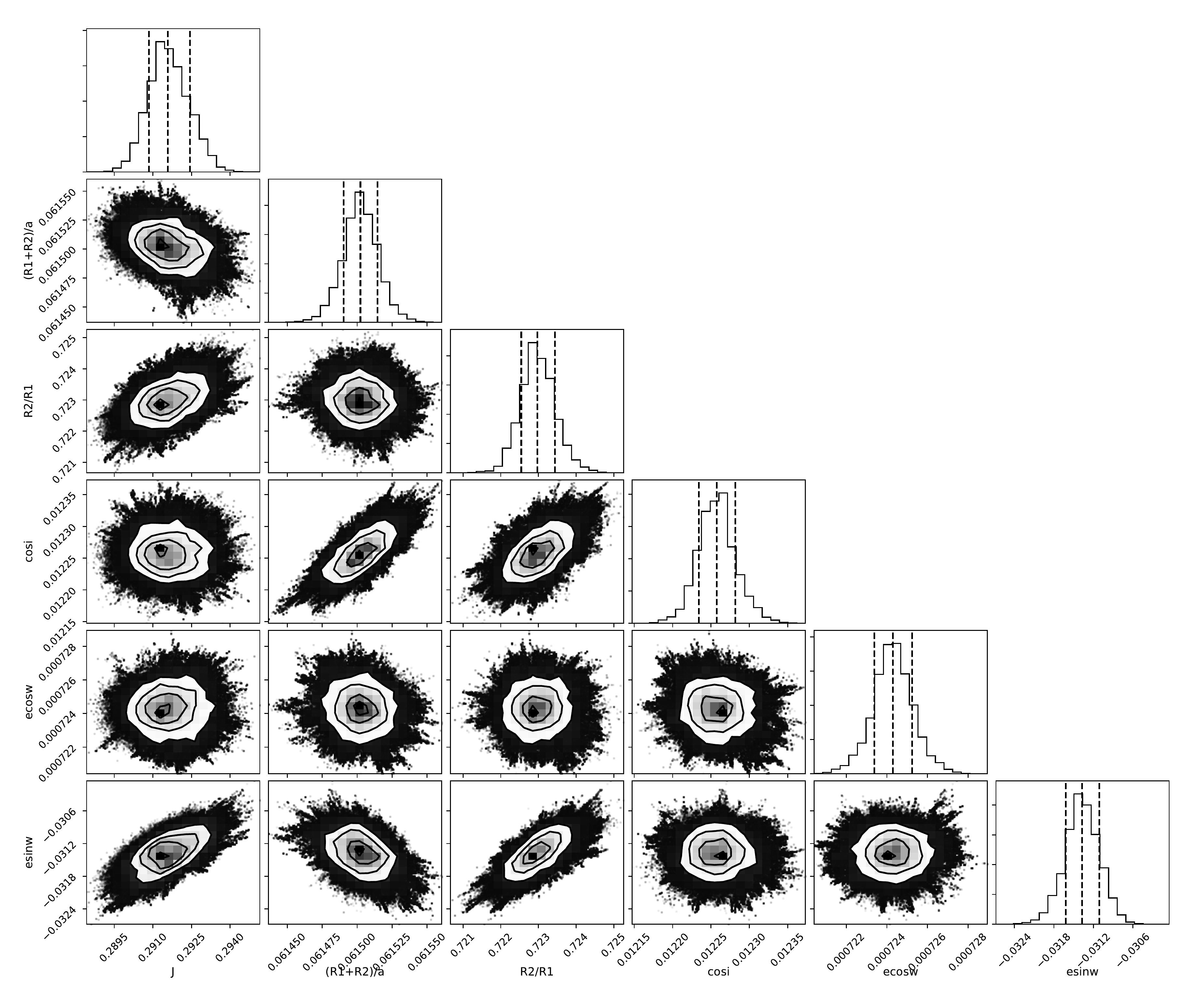}
\caption{Triangle plot of KIC 9821078 from the light curve fit. The histogram and the contour plots show density of MCMC. The dashed lines in the histogram mark 16$^{th}$, 50$^{th}$, and 84$^{th}$ percentiles of the samples in the marginalized distributions. 
See Table \ref{modelparm} for descriptions of the fitted parameters.}
\label{9821078_LC_corner}
\end{figure*}

\begin{figure*}
\centering
\includegraphics[width=7.0in]{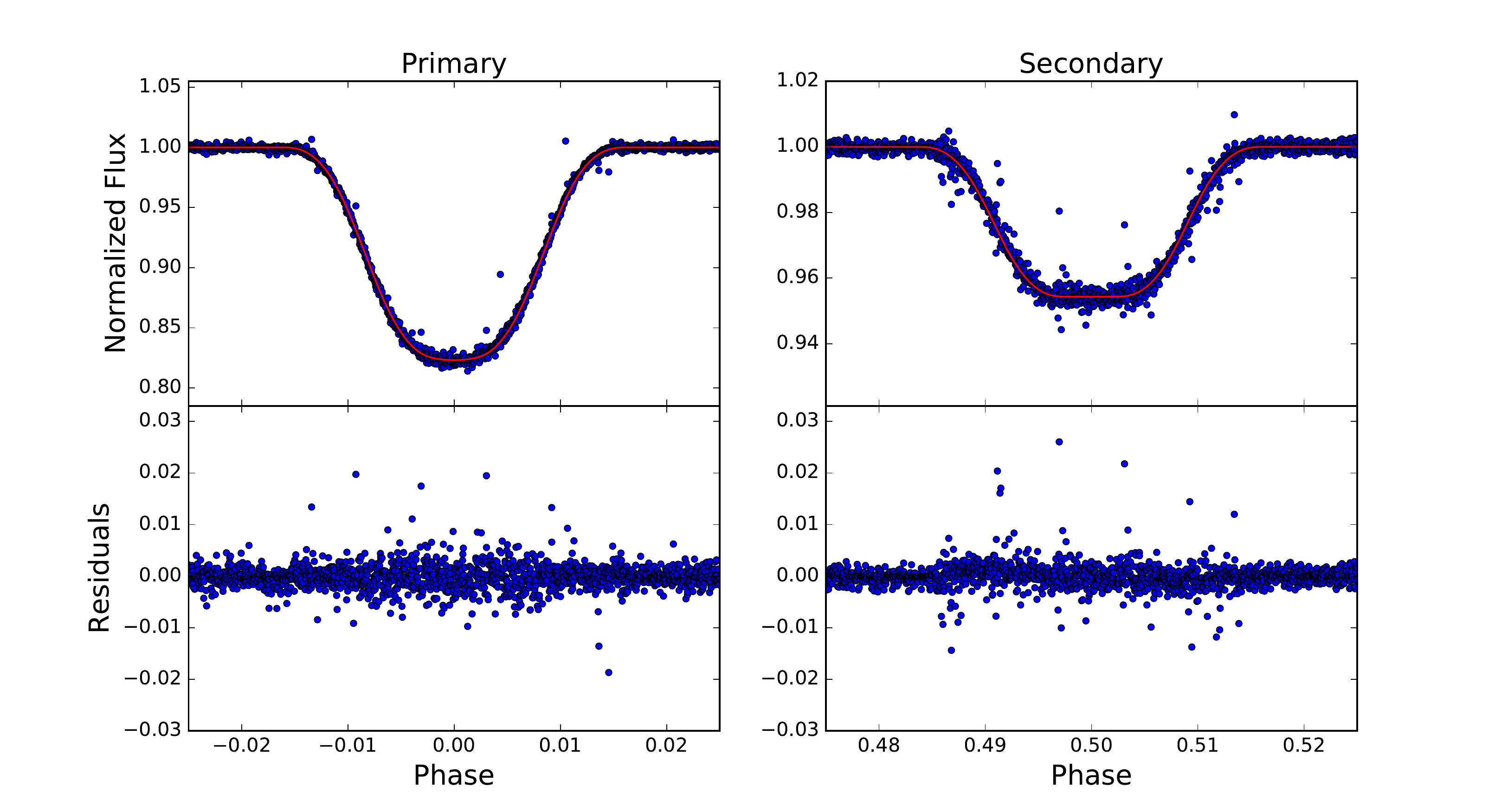}
\caption{Zoom-in of the phase-folded primary and secondary eclipses of KIC 7605600. The top panels show detrended and phase folded {\it Kepler} data with their best fit and the bottom panels show the residuals. We ascribe the scatter in the residuals to spot crossing events.}
\label{7605600_model_fit}
\end{figure*}

\begin{figure*}
\centering
\includegraphics[width=0.85\linewidth]{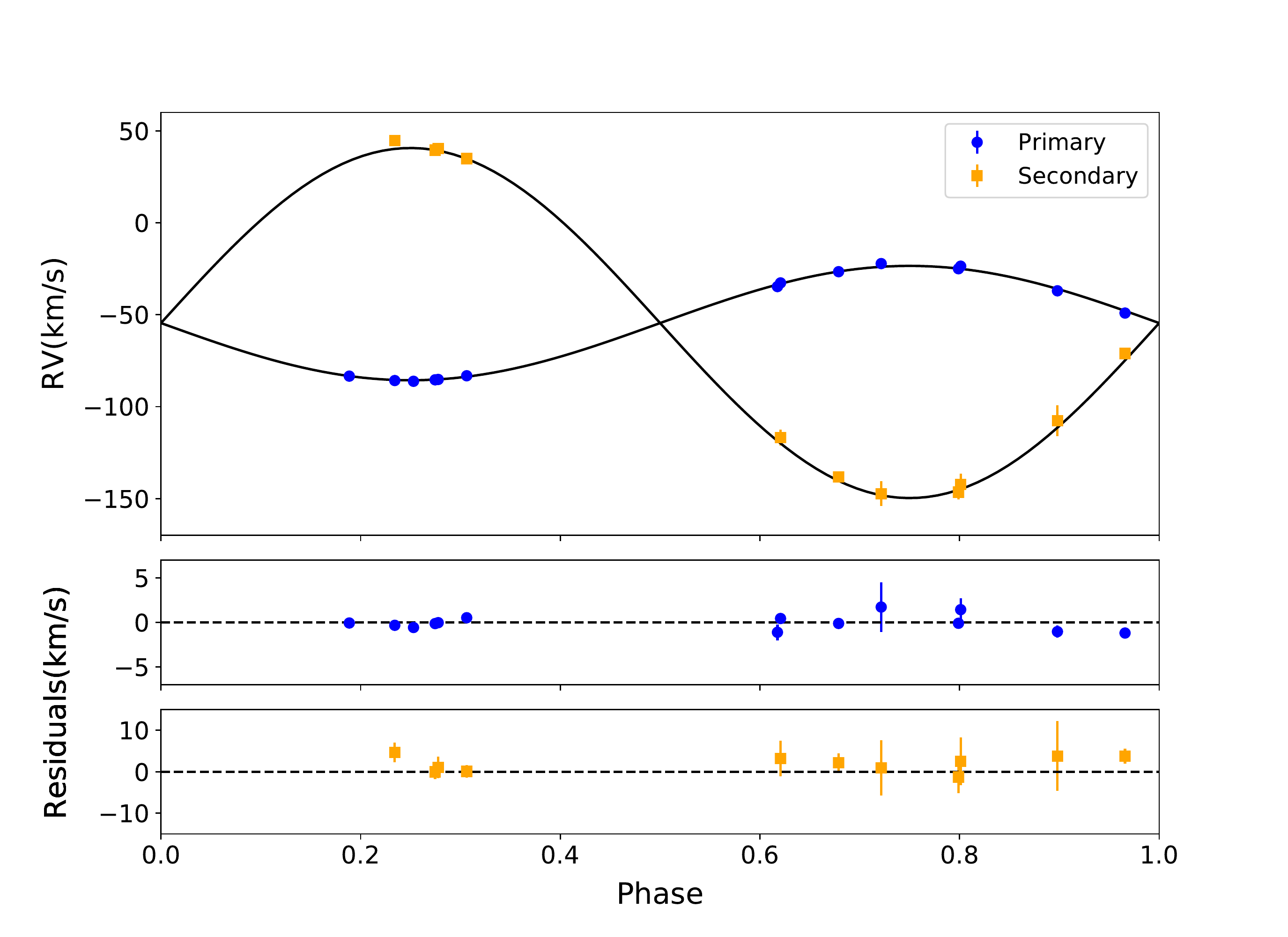}
\caption{Best-fit to the radial velocity data of KIC 7605600. In blue and in orange are the radial velocity data of the primary and the secondary, respectively. The solid black line is the analytically calculated best-fit model with IGRINS, iSHELL, and NIRSPEC data. The bottom two panels show the residuals for each component with their corresponding colors. The calculated radial velocity semi-amplitudes are $K_1 = 31.1 \pm 0.1 \,km\,s^{-1}$ for the primary and $K_2 = 95.8 \pm 0.5 \,km\,s^{-1}$ for the secondary using data from this work.} 
\label{7605600_rvfit}
\end{figure*}

\begin{figure*}
\centering
\includegraphics[width=0.75\linewidth]{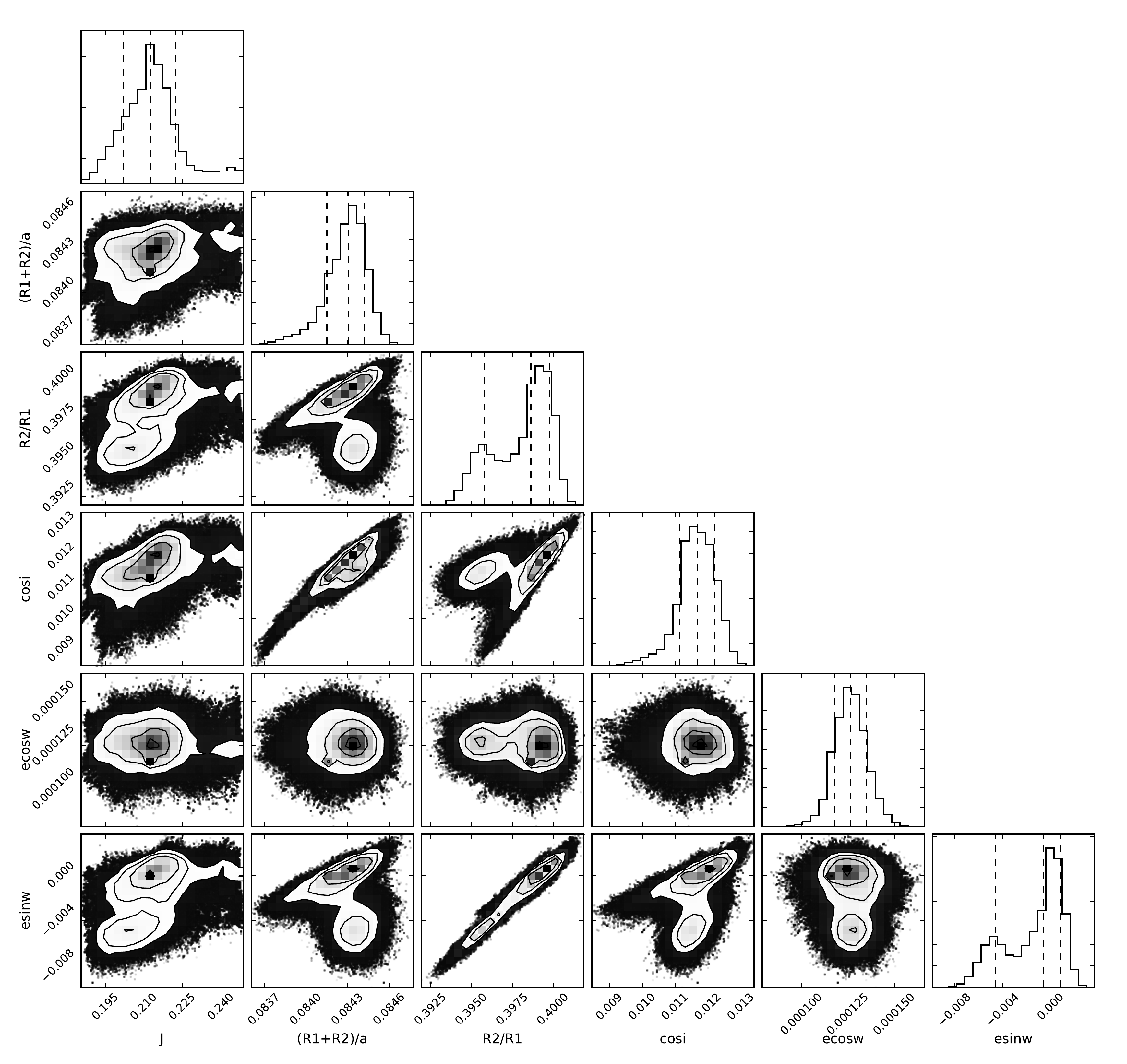}
\caption{Triangle plot of the light curve fit of KIC 7605600. The histogram and the contour plots show density of MCMC. The dashed lines in the histogram mark 16$^{th}$, 50$^{th}$, and 84$^{th}$ percentiles of the samples in the marginalized distributions. We investigated the bimodal posterior distribution of $R_2/R_1$ and e$\sin\omega$ and examined if either family of parameters results an inflated radius. The extracted masses and radii from each peak are consistent. We attribute the bimodal distribution to the high quality {\it Kepler} data where otherwise would be seen as unimodal. 
See Table \ref{modelparm} for descriptions of the fitted parameters.}
\label{7605600_LC_corner}
\end{figure*}

Figure \ref{11922782_model_fit}, \ref{9821078_model_fit}, and \ref{7605600_model_fit} show the zoom-in of the primary and the secondary eclipses of KIC 11922782, KIC 9821078, and KIC 7605600, respectively. The top panels show detrended and phase-folded {\it Kepler} data (in blue) with their best-fit models (in red) that we obtained using \texttt{eb}, and the bottom panels show the residuals. \\
\indent Figure \ref{11922672_LC_corner}, \ref{9821078_LC_corner}, and \ref{7605600_LC_corner} show the triangle plots of KIC 11922782, KIC 9821078, and KIC 7605600, respectively, from the MCMC run. We report the most probable value by taking a single step in the chains with the maximum likelihood. For the parameters with symmetric posterior distribution, We took the standard deviation of the chains to report as uncertainties. For the parameters with asymmetric posterior distributions, we took the difference between the values of the maximum likelihood and the 34.1$^{th}$ percentile around the most probable value and reported as asymmetric uncertainties.\\
\indent Figure \ref{11922782_rvfit}, \ref{9821078_rvfit}, and \ref{7605600_rvfit} show the radial velocity data of each component of KIC 11922782, KIC 9821078, and KIC 7605600, respectively. In the top panel, in blue and in orange are the radial velocity data of the primary and the secondary, respectively, and in black is the analytically calculated best-fit model. The bottom two panels show the residuals for each component with their corresponding colors. \\
\indent Table \ref{11922782_fitted_parameters}, \ref{9821078_fitted_parameters}, and \ref{7605600_fitted_parameters} show the fitted and the calculated parameters from the {\it Kepler} long-cadence and the SB2 radial velocity data fitting. For KIC 11922782, we measured $M_1 = 1.06 \pm 0.03 M_{\sun}$ and $R_1 = 1.53 \pm 0.02 R_{\sun}$ for the primary and $M_2 = 0.83 \pm 0.03 M_{\sun}$ and $R_2 = 0.88 \pm 0.01 R_{\sun}$ for the secondary. For KIC 9821078, we measured $M_1 = 0.67 \pm 0.01 M_{\sun}$ and $R_1 = 0.662 \pm 0.001 R_{\sun}$ for the primary and $M_2 = 0.52 \pm 0.01 M_{\sun}$ and $R_2 = 0.478 \pm 0.001 R_{\sun}$ for the secondary. Our measurements are consistent with the values reported by \citet[][]{Helminiak2017} and \citet[][]{Devor2008b}, respectively. For KIC 7605600, we measured $M_1 = 0.53 \pm 0.02 M_{\sun}$ and $R_1 = 0.501^{+0.001}_{-0.002} R_{\sun}$ for the primary and $M_2 = 0.17 \pm 0.01 M_{\sun}$ and $R_2 = 0.199^{+0.001}_{-0.002} R_{\sun}$ for the secondary. The secondary M dwarf component is fully-convective. \\

\begin{table*}
\begin{center}
\caption{Measured radial velocities for the primary and the secondary stars of KIC 11922782}
\begin{tabular}{@{}l c c c c c}
\hline
BJD & $V_1$ ($\,km\,s^{-1}$) & $\sigma_1$ ($\,km\,s^{-1}$) & $V_2$ ($\,km\,s^{-1}$) & $\sigma_2$ ($\,km\,s^{-1}$) & Instrument\\
\hline
2458387.74443477	&	25.5	&	0.6	&	-128.0	&	1.0	&	IGRINS	\\
2458388.62615568	&	-73.4	&	1.2	&	-0.4	&	1.2	&	IGRINS	\\
2458389.72504883	&	-91.2	&	0.7	&	19.2	&	1.6	&	IGRINS	\\
2458391.59712393	&	-2.7	&	0.7	&	-90.7	&	1.5	&	IGRINS	\\
2458407.64299461	&	-46.2	&	0.7	&	-	&	-	&	IGRINS	\\
2458416.66257339	&	-63.1	&	0.7	&	-9.2	&	0.4	&	IGRINS	\\
2458417.70374716	&	-104.4	&	0.8	&	39.5	&	1.2	&	IGRINS	\\
2458419.61101214	&	6.6	&	1.0	&	-101.2	&	1.2	&	IGRINS	\\
\hline
\label{11922782_rv}
\end{tabular}
\end{center}
\end{table*}

\begin{table*}
\begin{center}
\caption{Measured radial velocities for the primary and the secondary stars of KIC 9821078}
\begin{tabular}{@{}l c c c c c}
\hline
BJD & $V_1$ ($\,km\,s^{-1}$) & $\sigma_1$ ($\,km\,s^{-1}$) & $V_2$ ($\,km\,s^{-1}$) & $\sigma_2$ ($\,km\,s^{-1}$) & Instrument\\
\hline
2458388.67536168	&	-59.0	&	0.2	&	24.3	&	1.9	&	IGRINS	\\
2458389.64613792	&	-71.6	&	0.3	&	38.8	&	0.4	&	IGRINS	\\
2458417.71863644	&	-2.8	&	0.3	&	-49.9	&	0.7	&	IGRINS	\\
2458022.73752589	&	23.0	&	0.3	&	-82.8	&	1.0	&	IGRINS	\\
2453930.93654263	&	-72.4	&	0.7	&	36.7	&	0.8	&	NIRSPEC	\\
2453946.89327969	&	-67.2	&	1.1	&	32.4	&	1.7	&	NIRSPEC	\\
2453948.91693192	&	-46.5	&	0.4	&	3.2	&	0.5	&	NIRSPEC	\\
2454312.80698283	&	3.7	&	0.3	&	-60.0	&	0.6	&	NIRSPEC	\\
2458015.87697538	&	4.7	&	0.7	&	-64.8	&	3.2	&	iSHELL	\\
2458271.06522296	&	-65.9	&	0.8	&	-	&	-	&	iSHELL	\\
2458272.02832339	&	-73.8	&	2.6	&	-	&	-	&	iSHELL	\\
\hline
\label{9821078_rv}
\end{tabular}
\end{center}
\end{table*}

\begin{table*}
\begin{center}
\caption{Measured radial velocities for the primary and the secondary stars of KIC 7605600}
\begin{tabular}{@{}l c c c c c}
\hline
BJD & $V_1$ ($\,km\,s^{-1}$) & $\sigma_1$ ($\,km\,s^{-1}$) & $V_2$ ($\,km\,s^{-1}$) & $\sigma_2$ ($\,km\,s^{-1}$) & Instrument\\
\hline
2458020.6993352	&	-85.2	&	0.2	&	40.3	&	2.6	&	IGRINS	\\
2458058.6217044	&	-26.6	&	0.3	&	-138.2	&	2.2	&	IGRINS	\\
2458059.5761697	&	-49.1	&	0.2	&	-71.1	&	1.8	&	IGRINS	\\
2458060.6034628	&	-85.4	&	0.3	&	39.5	&	1.8	&	IGRINS	\\
2458389.7621102	&	-85.8	&	0.2	&	44.8	&	2.4	&	IGRINS	\\
2458391.6404928	&	-25.0	&	0.3	&	-146.5	&	3.9	&	IGRINS	\\
2458416.6113692	&	-83.2	&	0.2	&	34.9	&	1.5	&	IGRINS	\\
2458417.6571416	&	-32.7	&	0.3	&	-116.8	&	4.3	&	IGRINS	\\
2456844.9677341	&	-23.6	&	1.3	&	-142.3	&	5.8	&	NIRSPEC	\\
2456851.9425227	&	-37.0	&	0.7	&	-107.6	&	8.4	&	NIRSPEC	\\
2456858.0076913	&	-22.2	&	2.8	&	-147.3	&	6.7	&	NIRSPEC	\\
2458011.8511681	&	-34.7	&	0.9	&	-	&	-	&	iSHELL	\\
2458013.7506726	&	-83.4	&	0.4	&	-	&	-	&	iSHELL	\\
2458270.0815832	&	-86.2	&	0.5	&	-	&	-	&	iSHELL	\\
\hline
\label{7605600_rv}
\end{tabular}
\end{center}
\end{table*}

%%%%%%%%%%%%%%%%%%%%%%%%%%%%%%%%%%%%%%%%%%%%%%%%%%%%%%
% KIC 11922782

\begin{table*}
\begin{center}
\caption{Parameters for KIC 11922782 (this work).}
\begin{tabular}{l c c}
\hline
Fitted in Light Curve Analysis& Primary & Secondary\\
\hline
$J$ & \multicolumn{2}{c}{0.4888 $\pm$ 0.0011} \\
$(R_1 + R_2)/a$ & \multicolumn{2}{c}{0.2011 $\pm$ 0.0002} \\
$R_2/R_1$ & \multicolumn{2}{c}{0.5776 $\pm$ 0.0012} \\
$\cos{i}$ & \multicolumn{2}{c}{0.08389 $\pm$ 0.0003} \\
$P$ (days) & \multicolumn{2}{c}{3.512934029$\pm$ 0.000000003} \\
$T_0$ (BJD) & \multicolumn{2}{c}{2454956.2478567 $\pm$ 0.0000007} \\
$e \cos{\omega}$ & \multicolumn{2}{c}{0.0000683$\pm$ 0.0000004} \\
$e \sin{\omega}$ & \multicolumn{2}{c}{-0.0078$\pm$0.0008} \\
$L3$ & \multicolumn{2}{c}{0.0 (fixed)}\\
$LDLIN$ $K_P$ &0.230 $\pm$ 0.081 & 0.328 $\pm$ 0.087 \\
$LDNON$ $K_P$ & 0.770 $\pm$ 0.088 & 0.552 $\pm$ 0.085\\
\hline
Fitted in Radial Velocity Analysis& Primary & Secondary\\
\hline
$\gamma$ ($\,km\,s^{-1}$) & \multicolumn{2}{c}{-41.4$\pm$ 0.1} \\
q & \multicolumn{2}{c}{0.783 $\pm $ 0.001}\\
$K_{tot}$ ($\,km\,s^{-1}$) & \multicolumn{2}{c}{172.1 $\pm$ 0.1}\\
\hline
Calculated & Primary & Secondary \\
\hline
$e$ & \multicolumn{2}{c}{0.0078 $\pm$ 0.0003}\\
$i$ ($^\circ$) & \multicolumn{2}{c}{85.19 $\pm$ 0.01}  \\
$a_{\rm tot}$ (R$_{\sun}$) & \multicolumn{2}{c}{12.02 $\pm$ 0.15} \\
$K$ ($\,km\,s^{-1}$) & 75.6 $\pm$ 0.1 & 96.6 $\pm$ 0.1 \\
$M$ ($\rm M_{\sun}$) & 1.06 $\pm$ 0.03 & 0.83 $\pm$ 0.03 \\
$R$ ($\rm R_{\sun}$) & 1.53 $\pm$ 0.02 & 0.88 $\pm$ 0.01 \\
\label{11922782_fitted_parameters}
\end{tabular}
\end{center}
\end{table*}

%%%%%%%%%%%%%%%%%%%%%%%%%%%%%%%%%%%%%%%%%%%%%%%%%%%%%%
% KIC 9821078

\begin{table*}
\begin{center}
\caption{Parameters for KIC 9821078 (this work).}
\begin{tabular}{@{}l c c}
\hline
Fitted in Light Curve Analysis& Primary & Secondary\\
\hline
$J$ & \multicolumn{2}{c}{0.2916 $\pm$ 0.0006} \\
$(R_1 + R_2)/a$ & \multicolumn{2}{c}{0.061502 $\pm$ 0.000001} \\
$R_2/R_1$ & \multicolumn{2}{c}{0.7230 $\pm$ 0.0003} \\
$\cos{i}$ & \multicolumn{2}{c}{0.01226 $\pm$ 0.00001} \\
$P$ (days) & \multicolumn{2}{c}{8.4294382$\pm$ 0.0000002}\\
$T_0$ (BJD) & \multicolumn{2}{c}{2454965.2922462 $\pm$ 0.0000194} \\
$e \cos{\omega}$ & \multicolumn{2}{c}{0.0007243$\pm$ 0.0000007} \\
$e \sin{\omega}$ & \multicolumn{2}{c}{-0.0314$\pm$0.0002} \\
$L3$ & \multicolumn{2}{c}{0.0 (fixed)}\\
$LDLIN$ $K_P$ & $0.696 ^{+0.304}_{-0.437}$ & $0.138^{+0.278}_{-0.138}$ \\
$LDNON$ $K_P$ & $ 0.0001^{+0.0059}_{-0.0002} $ & $0.0018^{+0.0599}_{-0.0013}$\\

\hline
Fitted in Radial Velocity Analysis& Primary & Secondary\\
\hline
$\gamma$ ($\,km\,s^{-1}$) & \multicolumn{2}{c}{-23.8$\pm$ 0.1} \\
q & \multicolumn{2}{c}{0.777 $\pm $ 0.002}\\
$K_{tot}$ ($\,km\,s^{-1}$) & \multicolumn{2}{c}{111.6 $\pm$ 0.1}\\

\hline
Calculated Parameters & Primary & Secondary \\
\hline
$e$ & \multicolumn{2}{c}{0.0314 $\pm$ 0.0002}\\
$i$ ($^\circ$) & \multicolumn{2}{c}{89.297 $\pm$ 0.001}  \\
$a_{\rm tot}$ (R$_{\sun}$) & \multicolumn{2}{c}{18.516 $\pm$ 0.034} \\
$K$ ($\,km\,s^{-1}$) & 48.8 $\pm$ 0.1 & 62.8 $\pm$ 0.1 \\
$M$ ($\rm M_{\sun}$) & 0.67 $\pm$ 0.01 & 0.52 $\pm$ 0.01 \\
$R$ ($\rm R_{\sun}$) & 0.662 $\pm$ 0.001 & 0.478 $\pm$ 0.001 \\
\label{9821078_fitted_parameters}
\end{tabular}
\end{center}
\end{table*}

%%%%%%%%%%%%%%%%%%%%%%%%%%%%%%%%%%%%%%%%%%%%%%%%%%%%%%
% KIC 7605600

\begin{table*}
\begin{center}
\caption{Parameters for KIC 7605600 (this work).}
\begin{tabular}{@{}l c c}
\hline
Fitted in Light Curve Analysis& Primary & Secondary\\
\hline
$J$ & \multicolumn{2}{c}{0.2012 $\pm$ 0.0119} \\
$(R_1 + R_2)/a$ & \multicolumn{2}{c}{0.0840 $\pm$ 0.0001} \\
$R_2/R_1$ & \multicolumn{2}{c}{$ 0.3980^{+0.0014}_{-0.0032} $}\\
$\cos{i}$ & \multicolumn{2}{c}{0.01083 $\pm$ 0.00059} \\
$P$ (days) & \multicolumn{2}{c}{3.32619385$\pm$ 0.00000005}\\
$T_0$ (BJD) & \multicolumn{2}{c}{2455006.2441023$\pm$ 0.000013} \\
$e \cos{\omega}$ & \multicolumn{2}{c}{0.000133$\pm$ 0.000009} \\
$e \sin{\omega}$ & \multicolumn{2}{c}{$ -0.0013^{+0.0017}_{-0.0043}$} \\
$L3$ & \multicolumn{2}{c}{0.0 (fixed)}\\
$LDLIN$ $K_P$ & $0.5427 ^{+0.232}_{-0.017}$ & $-0.391^{+0.348}_{-0.411}$ \\
$LDNON$ $K_P$ & $ 0.0116^{+0.148}_{-0.261} $ & $-0.606^{+0.539}_{-0.548}$\\
\hline
Fitted in Radial Velocity Analysis& Primary & Secondary\\
\hline
$\gamma$ ($\,km\,s^{-1}$) & \multicolumn{2}{c}{-54.6$\pm$ 0.1} \\
q & \multicolumn{2}{c}{0.32 $\pm $ 0.01}\\
$K_{tot}$ ($\,km\,s^{-1}$) & \multicolumn{2}{c}{126.9 $\pm$ 0.5}\\
\hline
Calculated & Primary & Secondary \\
\hline
$e$ & \multicolumn{2}{c}{$0.0013^{+0.0043}_{-0.0008}$}\\
$i$ ($^\circ$) & \multicolumn{2}{c}{89.38 $\pm$ 0.03}  \\
$a_{\rm tot}$ (R$_{\sun}$) & \multicolumn{2}{c}{8.86 $\pm$ 0.01} \\
$K$ ($\,km\,s^{-1}$) & 31.1 $\pm$ 0.1 & 95.8 $\pm$ 0.5 \\
$M$ ($\rm M_{\sun}$) & 0.53 $\pm$ 0.02 & 0.17 $\pm$ 0.01 \\
$R$ ($\rm R_{\sun}$) & $0.501 ^{+0.001}_{-0.002}$& $0.199 ^{+0.001}_{-0.002}$\\
\label{7605600_fitted_parameters}
\end{tabular}
\end{center}
\end{table*}

%%%%%%%%%%%%%%%%%%%%%%%%%%%%%%%%%%%%%%%%%%%%%%%%%%%%%%%%%%%%%%%%%%%%%%%%%%%%%%%%%%%%%%%%%%%
\section{Discussion}\label{Discussion}
\indent Following the same method as described in \citep[][]{Han2017}, our independent measurements for the two previously published systems, KIC 11922782 and KIC 9821078, are consistent with the literature. Among the 3 systems, KIC 9821078 and KIC 7605600 contain at least one M dwarf stars and we discuss the two systems in detail. 

\subsection{M dwarf SB2 EBs}
\subsubsection{KIC 9821078}
\begin{figure*}
\centering
\includegraphics[width=0.35\linewidth]{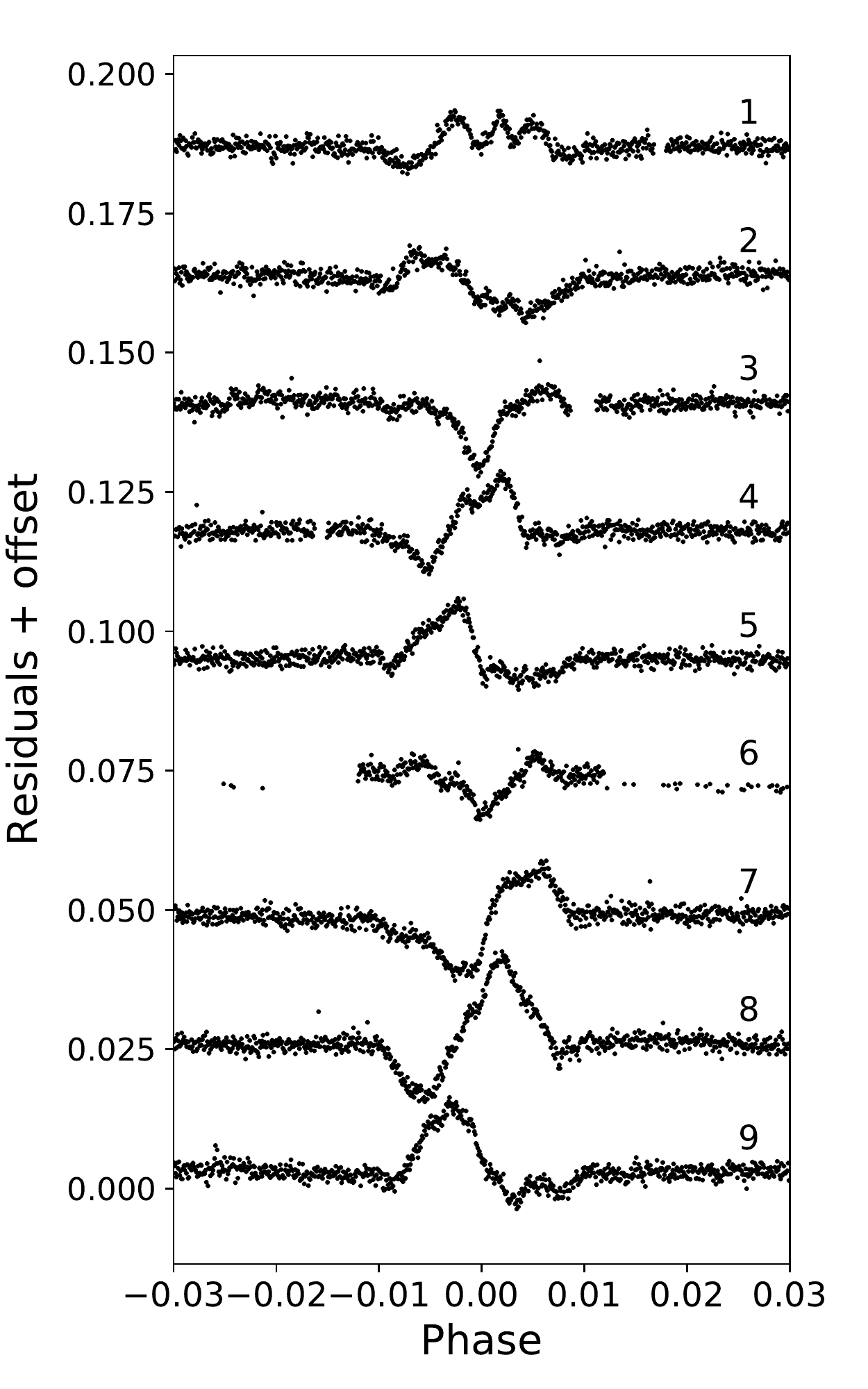}
\includegraphics[width=0.35\linewidth]{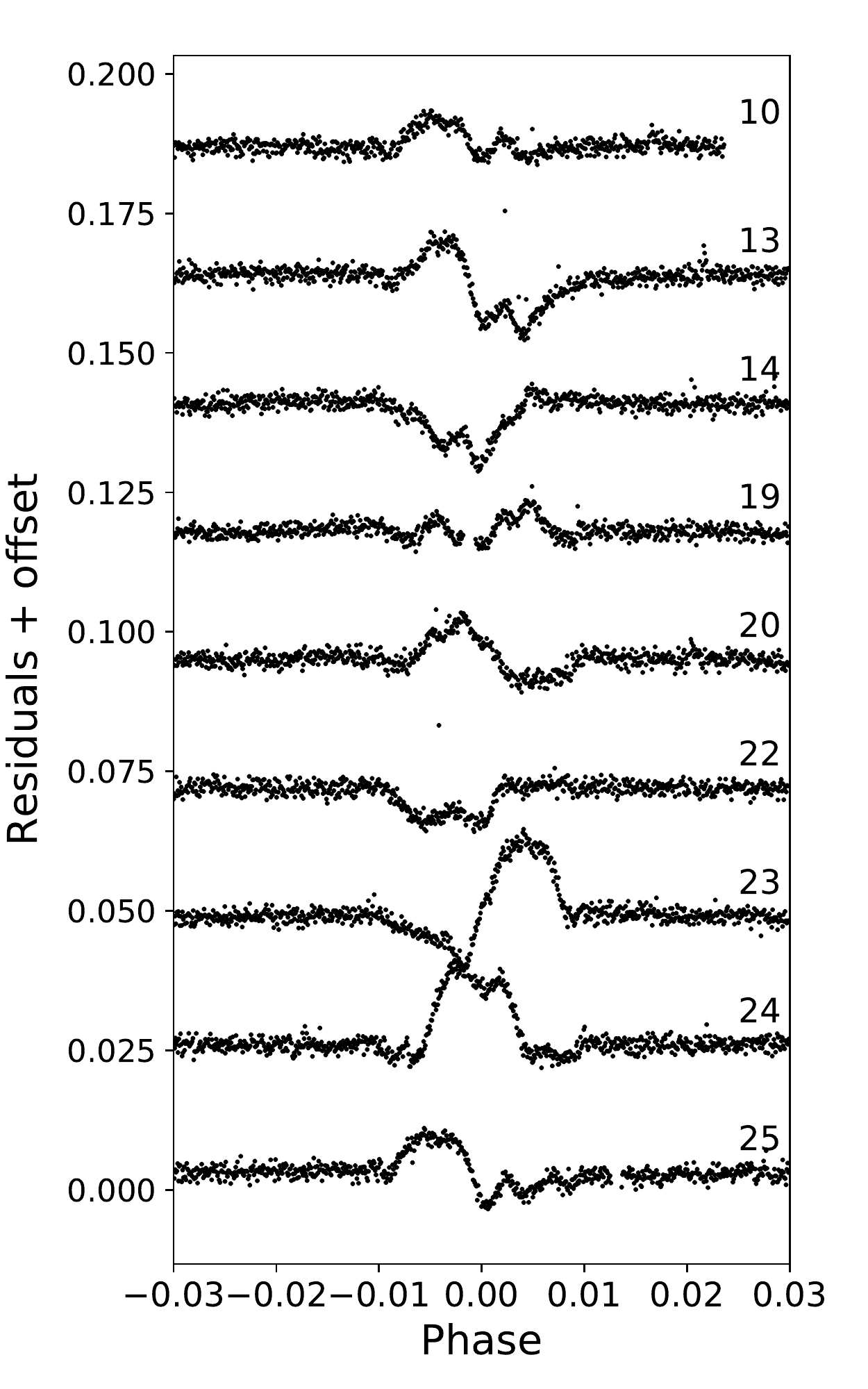}
\includegraphics[width=0.35\linewidth]{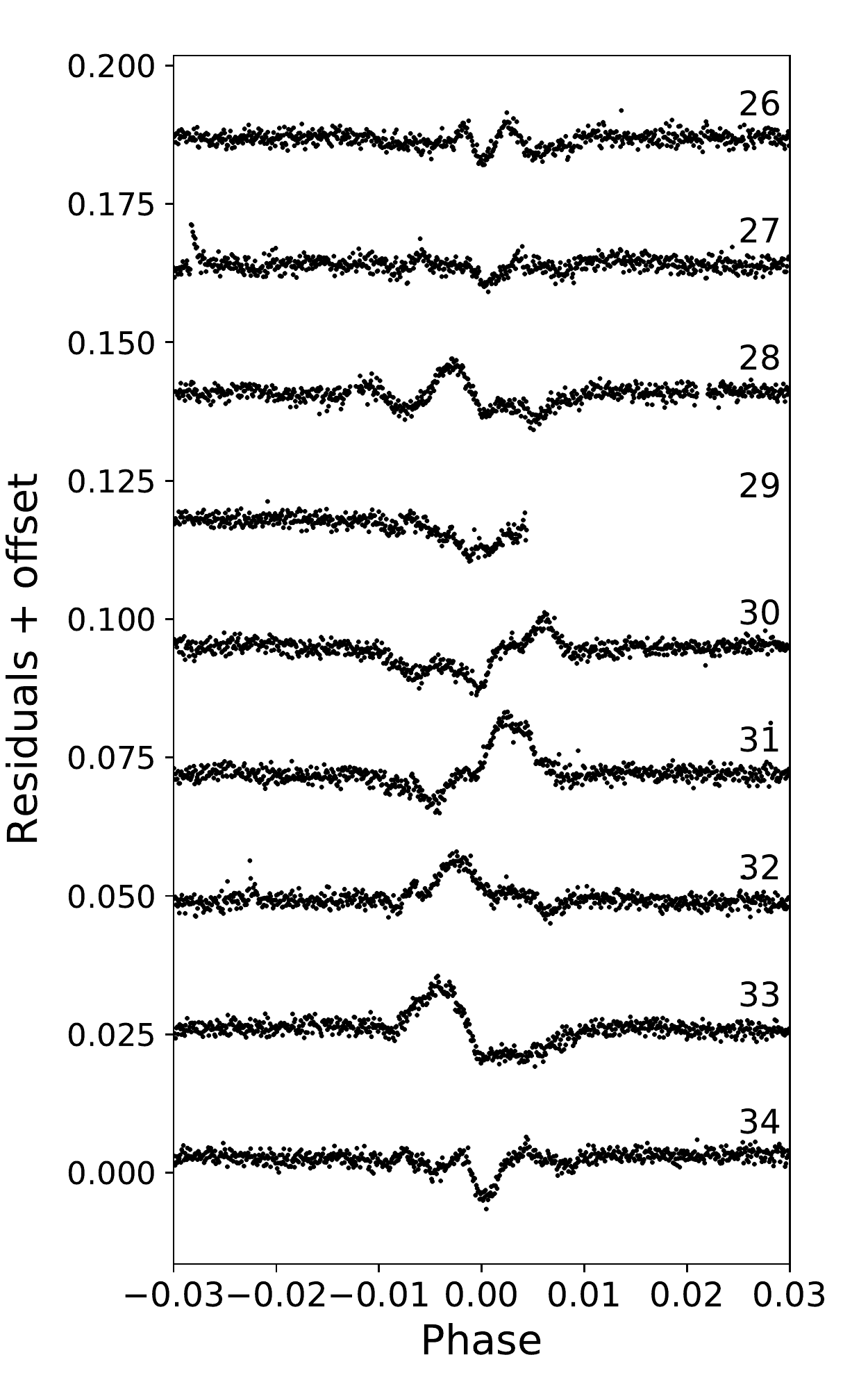}
\includegraphics[width=0.35\linewidth]{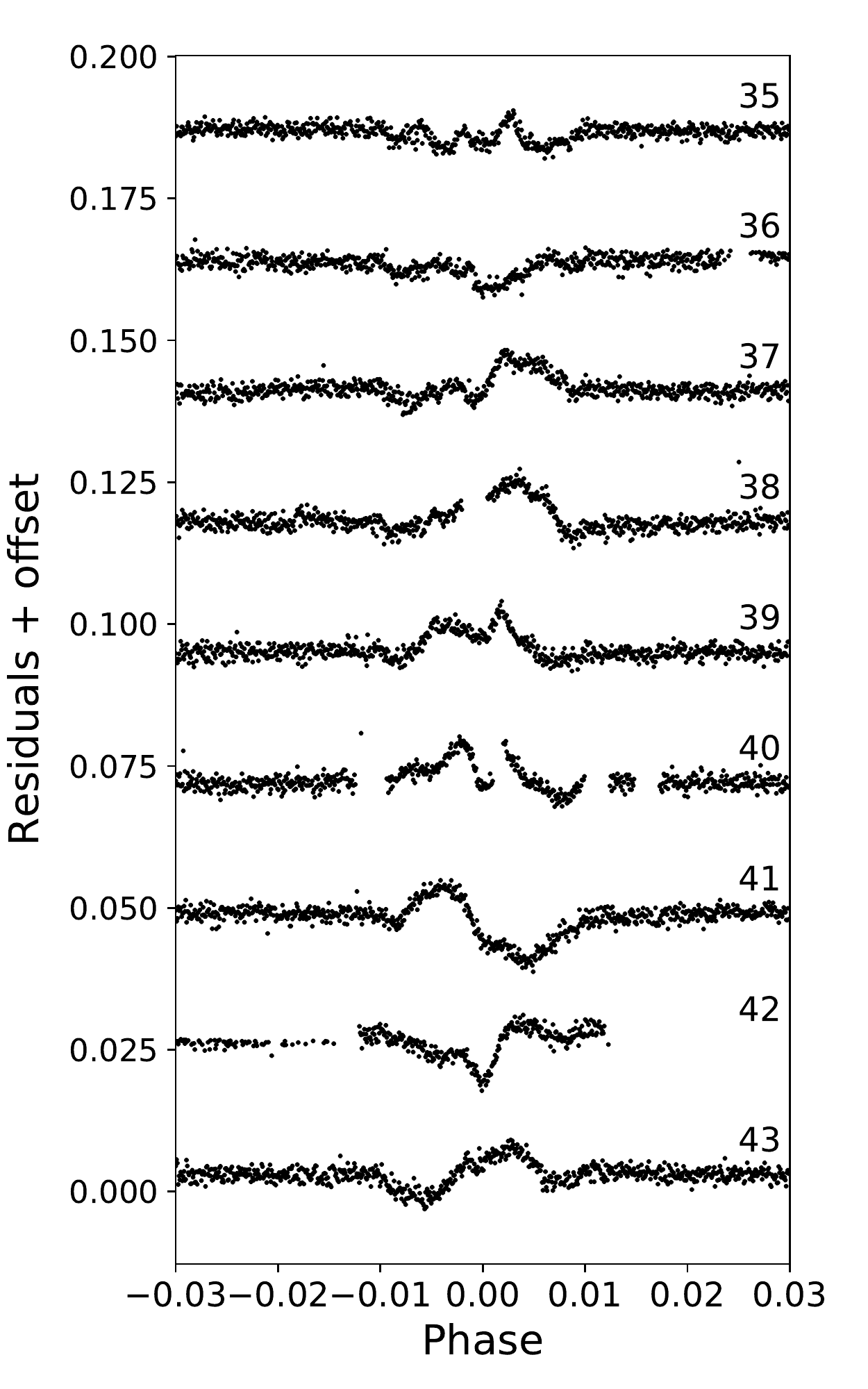}
\caption{Residual plots of the best-fit model and the {\it Kepler} short-cadence primary eclipse data of KIC 9821078. The eclipses are near 0.0 in phase. The eclipses are numbered sequentially with skipped numbers indicating eclipses missed by the {\it Kepler} spacecraft. From the 36 primary eclipses, it is evident that starspots were occulted during the eclipses and evolve over time. }
\label{9821078_spot_crossing_primary}
\end{figure*}

\begin{figure*}
\centering
\includegraphics[width=0.35\linewidth]{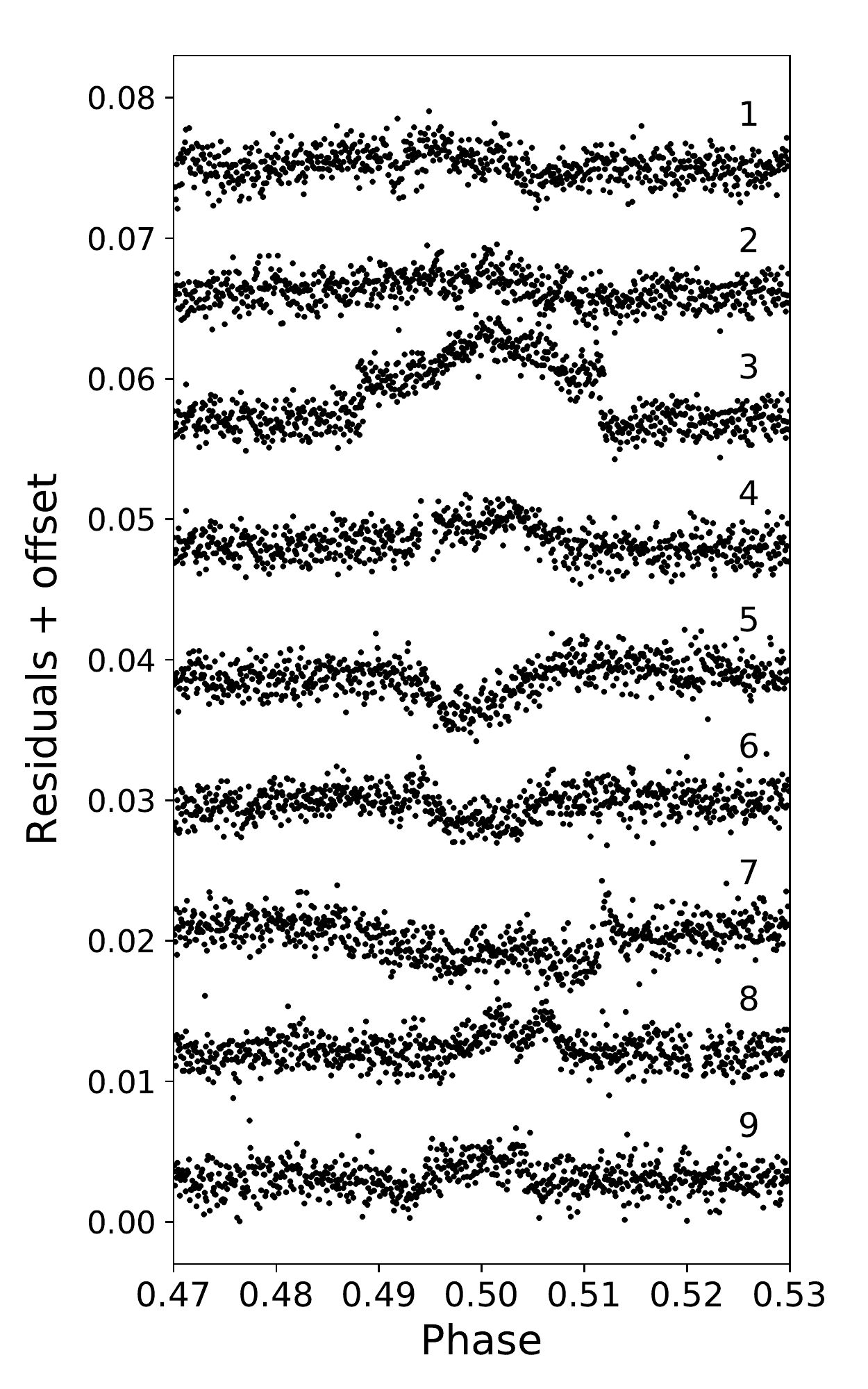}
\includegraphics[width=0.35\linewidth]{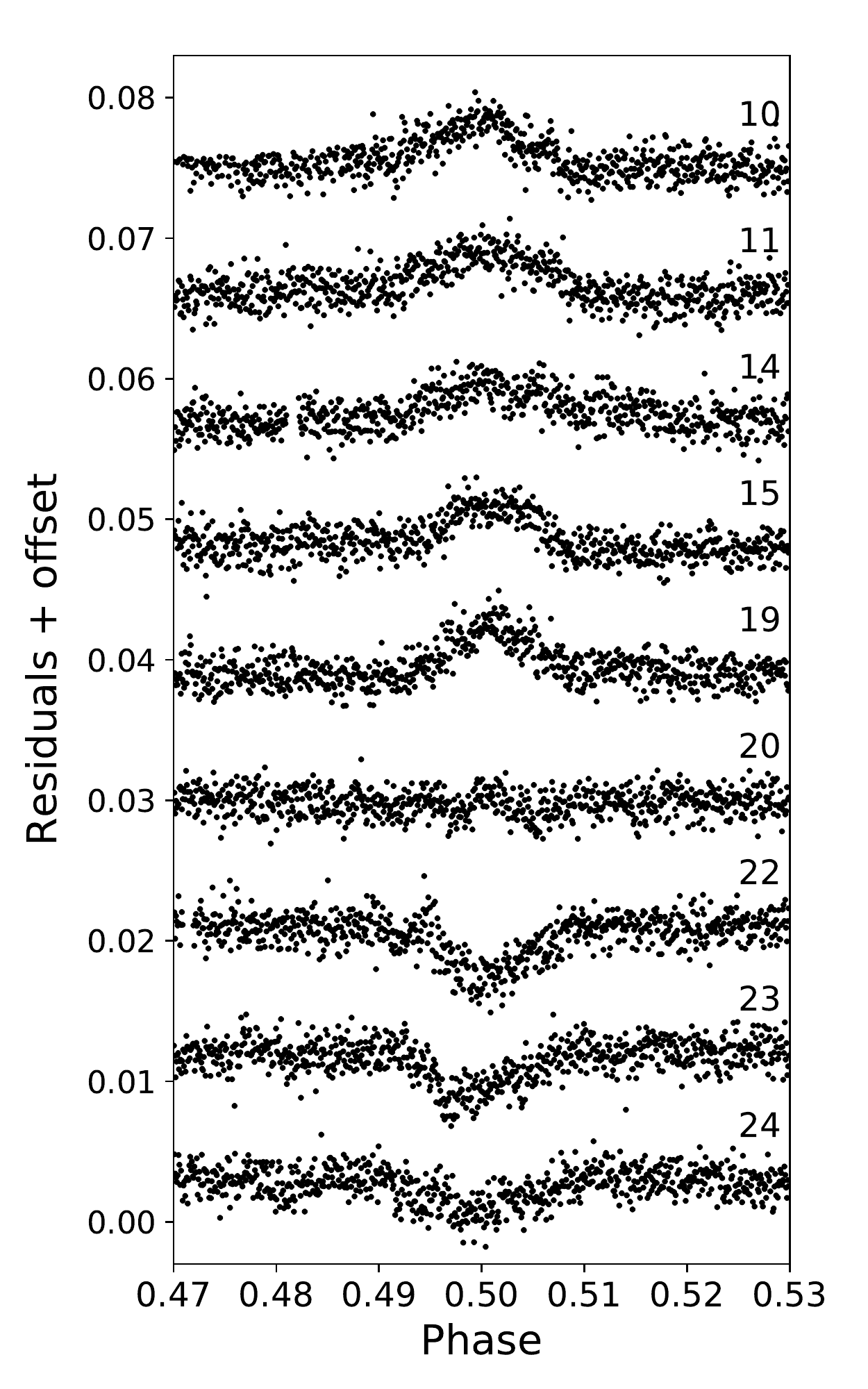}
\includegraphics[width=0.35\linewidth]{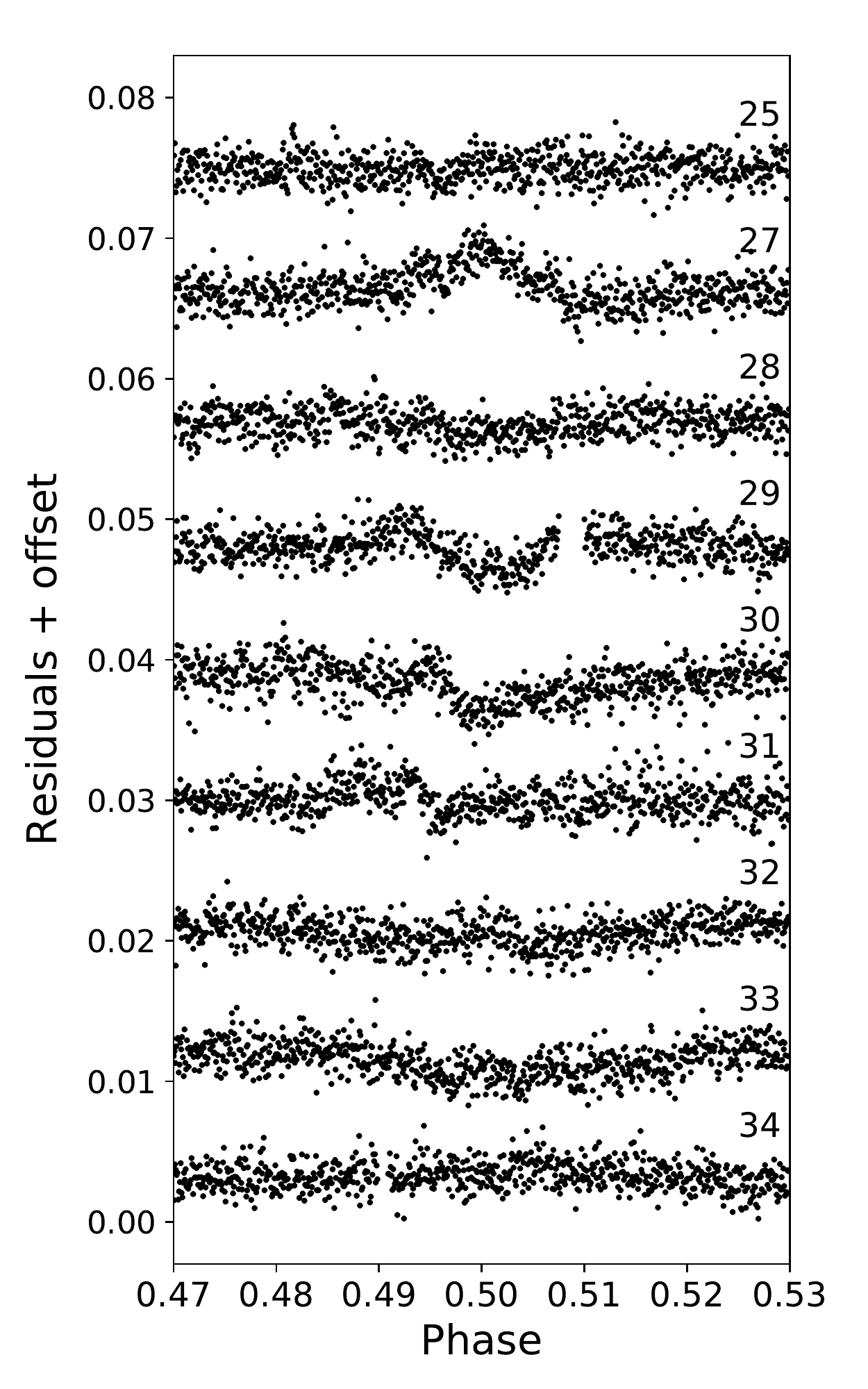}
\includegraphics[width=0.35\linewidth]{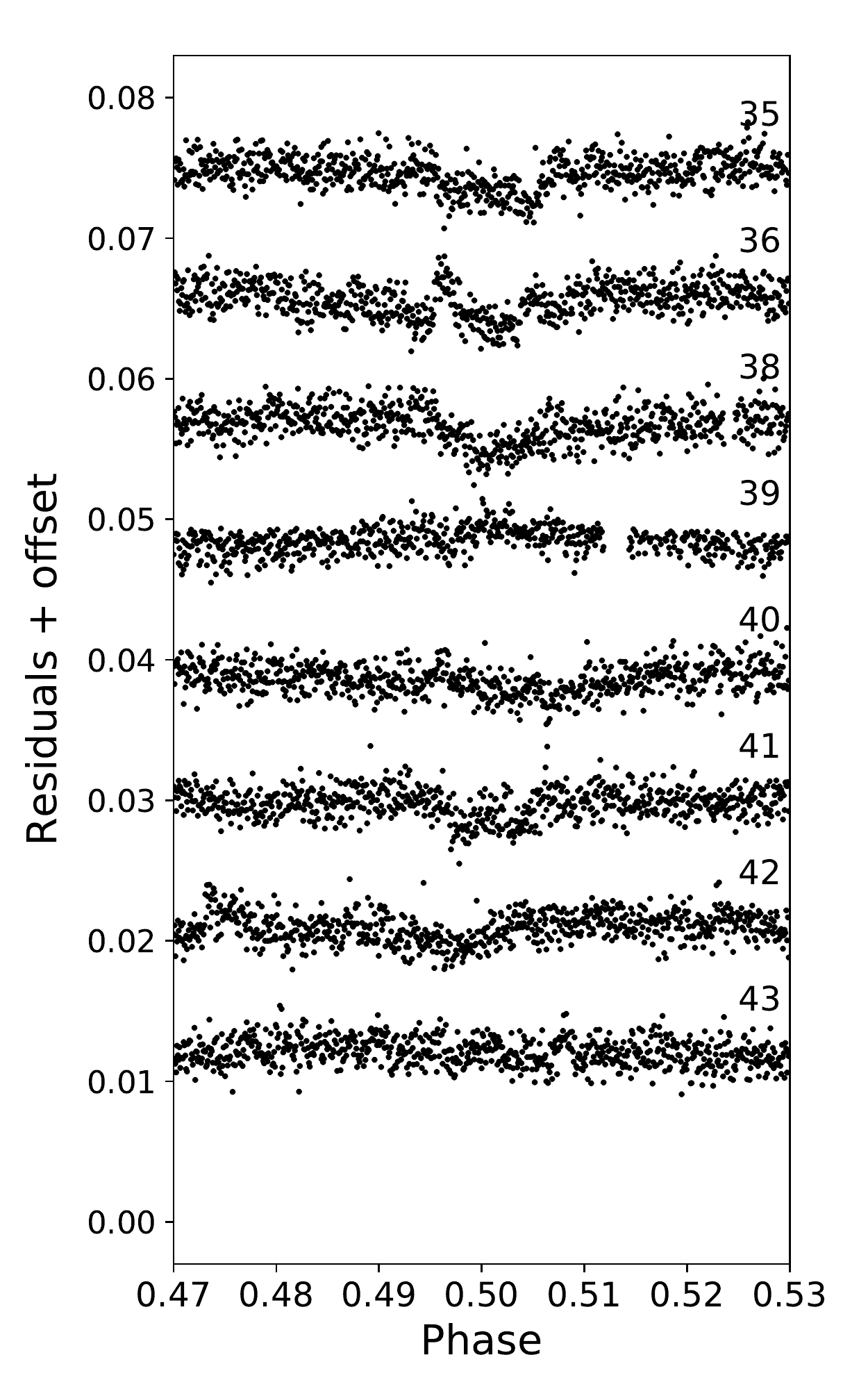}
\caption{Residual plots of the best-fit model and the {\it Kepler} short-cadence secondary eclipse data of KIC 9821078. The eclipses are near 0.5 in phase. The eclipses are numbered sequentially with skipped numbers indicating eclipses missed by the {\it Kepler} spacecraft. From the 36 secondary eclipses, although the amplitudes of the residuals are smaller than that of the primary eclipse, it is still shown that starspots were occulted during the secondary eclipses and evolve as well.}
\label{9821078_spot_crossing_secondary}
\end{figure*}

KIC 9821078 is an EB with a late-K dwarf primary and an early-M dwarf secondary, based on the measured masses. The distance to the system is $\sim$243 pc, measured by Gaia \citep[][]{GaiaDR2}. As shown in Figure \ref{kic9821078_sample_lc}, the short-cadence {\it Kepler} data show spot-crossings during both the primary and the secondary eclipses. Figure \ref{9821078_spot_crossing_primary} and \ref{9821078_spot_crossing_secondary} present the residuals of the best-fit model and the {\it Kepler} short-cadence data where the positive deviations from 0 indicate dark spots occulted during the eclipse. The eclipses are numbered sequentially with skipped numbers indicating eclipses missed by the {\it Kepler} spacecraft. Spot occultations can give estimates on the distribution of spots on the stellar surface, especially when the component stars have synchronous rotational periods with the orbital period since the same side of the stars are visible during the eclipses. By inspection, similar residual features are repeated (e.g. 7$^{th}$, 8$^{th}$, and 9$^{th}$ \& 23$^{rd}$, 24$^{th}$, and 25$^{th}$ in Figure \ref{9821078_spot_crossing_primary} and \ref{9821078_spot_crossing_secondary}) but their positions are slightly offset from each other. These features could be caused by synchronized stars with their spots evolving, differential stellar rotation, or slightly subsychronous rotation of stars with their spots evolving. We argue that the stars are synchronized with spots evolving over time, but given the scope of our work, we do not discuss the details on the spot evolution timescale in this paper. The light curve of KIC 9821078 shows $\sim$5\% rotational spot modulations and flares. These indicate that the component stars are magnetically active.  \\
\indent The eccentricity of the orbit is very small but non-zero. This is shown in Figure \ref{9821078_model_fit} where in the phase-folded light curve, the secondary mid-eclipse time slightly departs from 0.5 in orbital phase. Furthermore, the component stars of KIC 9821078 are tidally-locked and their rotation periods match the orbital period of the system. 

\subsubsection{KIC 7605600}

\indent KIC 7605600 is a newly measured M+M SB2 EB system. The system has a parallax of 6.26 $\mu$as, measured by Gaia \citep[][]{GaiaDR2}, hence the distance of $\sim$160 pc. Although no short-cadence {\it Kepler} data is available, the 8 quarters of long-cadence data provide ample coverage. The residuals from the best-fit model shown in Figure \ref{7605600_model_fit} shows the spot crossing events during both the primary and the secondary eclipses. \\
\indent The {\it Kepler} long-cadence data show the out-of-eclipse modulations. The possible causes of these modulations are star spots rotating in and out of the line-of-sight, reflected light from the other component, ellipsoidal variations, beaming effects, and gravity-darkening. For low-mass stars like KIC 7605600, gravity-darkening is negligible as predicted by von Zeipel Theorem \citep[][]{vonZeipel1924} where the effect is significant for stars with radiative envelopes. Given our best fit parameters, we computed three different light curve models, each containing an effect of the reflection, ellipsoidal variations, and beaming, respectively. Their signals in the out-of-eclipse portion of the light curve were negligible. Therefore, we  attribute the cause of modulation to star spots. \\
\indent The flat-bottomed secondary eclipse indicates a total eclipse where the secondary component is completely blocked by the primary component. From our best-fit model, the calculated secondary eclipse depth is $\sim$4.5\%, which indicates the contribution of the primary component to the total flux is $\sim$95.5\%. From the reported {\it Kepler} magnitude in MAST, we calculated the individual magnitude of the component stars in {\it Kepler} band, which are 14.94 and 18.24, respectively. Incorporating the parallax measured by Gaia, the absolute {\it Kepler} band magnitudes of the primary and the secondary are 8.92 and 12.22, respectively. Our fitting method does not fit for the effective temperatures and we purposefully report the central surface brightness ratio in the {\it Kepler} band, instead, to avoid any assumptions about metallicity. Determining the effective temperatures of the stars involves atmospheric models, which are known to disagree with spectroscopic observations, due to rich molecular lines in the spectra of low-mass stars \citep[][]{Veyette2016}. \\
\indent The amplitude of the out-of-eclipse modulation caused by spots is $\sim$3\%. This is comparable to the secondary eclipse depth and we conclude that the primary component is magnetically active. The magnetic activity of the primary star is also evident in Figure \ref{kic7605600_sample_lc} where the long-cadence {\it Kepler} data contains flares shortly before and after the secondary eclipses. \\
\indent We report the standard deviation of the MCMC chains as the uncertainty for all of the parameters with symmetric posterior distribution. For parameters with asymmetric posterior distribution (e.g. $R_2/R_1$, $e\sin\omega$, and the limb-darkening parameters), we took the difference between the values of the maximum likelihood and the 34.1$^{th}$ percentile around the maximum likelihood and reported as asymmetric uncertainties. We also note that the limb-darkening parameters were not well-constrained from our fit, as shown by their uncertainties. However, these uncertainties are folded in the uncertainties of the other extracted parameters. We investigated the bimodal posterior distribution of $R_2/R_1$ and e$\sin\omega$ to examine if either family of parameters result an inflated radius. We calculated the masses and radii using the corresponding chain for each peak in the bimodal posterior distribution of $R_2/R_1$ and $e\sin\omega$ and ensure the extracted masses and radii from each peak are consistent.\\
\indent Both the circularization timescale ($\tau_{circ}$) and the synchronization timescale ($\tau_{sync}$) can be used to infer the age of the system. These time scales are proportional to $\simeq(a/R_1)^8$ and $\simeq(a/R_1)^6$, respectively, for solar type stars. For fully convective stars, the synchronization timescale is suggested to be longer than the prediction from the theory of equilibrium tides \citep[][]{Gillen2017}. The rotation of the component stars synchronize with the orbital motion and the binary orbit is circularized. For KIC 7605600, these timescales are $\tau_{circ}$ $\sim$5.72 Gyr and $\tau_{sync}$ $\sim$21 Myr. Our analysis shows that both components have synchronous rotation that match with the orbital periods with a circular orbit. This imposes a lower limit on the age of the system, which is on the order of several Gyr. \\

\section{Conclusions}
\begin{figure*}
\centering
\includegraphics[width=0.75\linewidth]{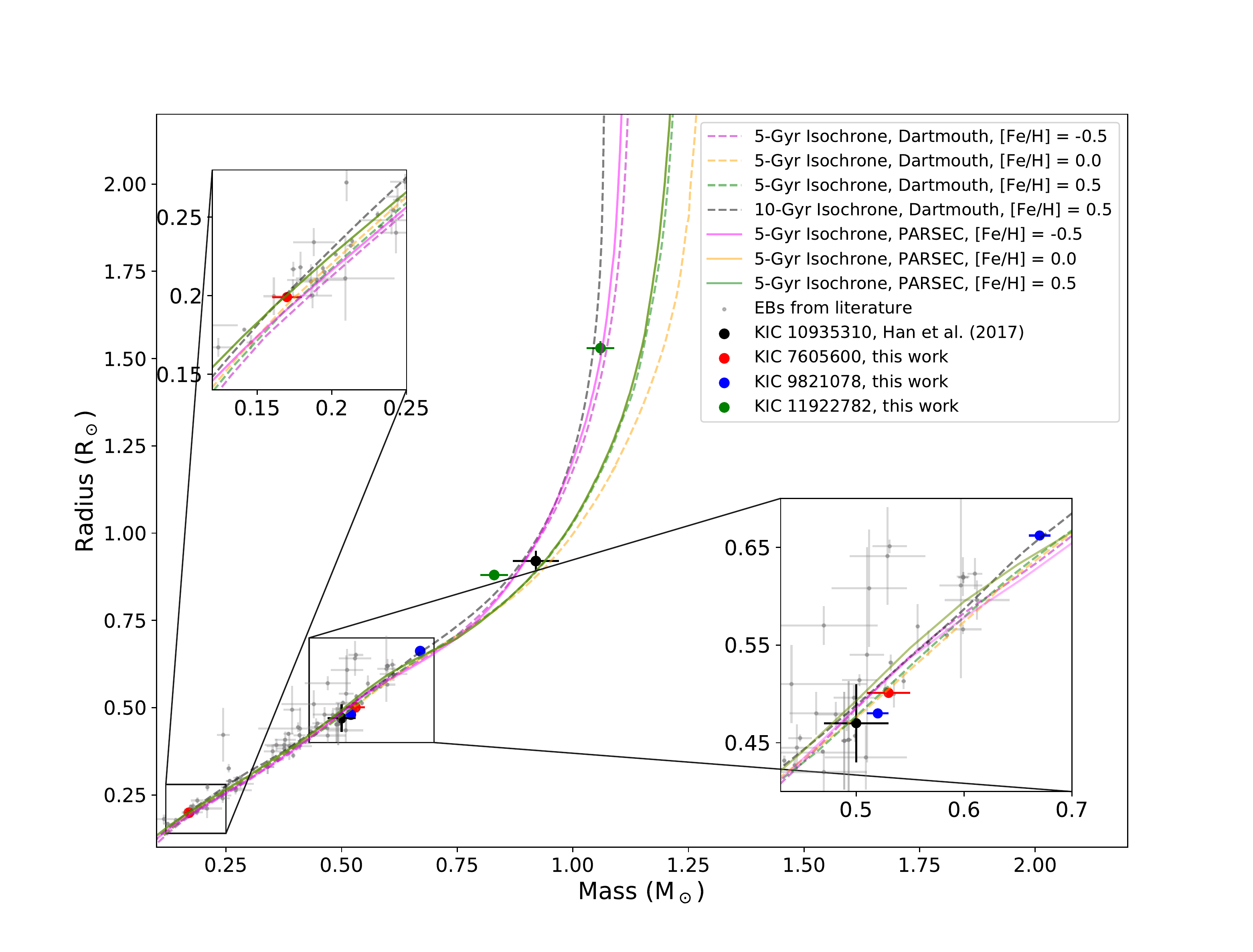}
\caption{Mass vs. radius for main-sequence low-mass EBs \citep[{\it gray circles}, see][for references]{Parsons2018} with our measurements in red, blue, green, and black circles. We include predictions for 5 and 10-Gyr-old stars from the Dartmouth evolutionary isochrones \citep{Dotter2008} as dashed lines and from the PARSEC evolutionary isochrones \citep[][]{PARSEC} as solid lines for three metallicities: [M/H] = -0.5, 0.0 and +0.5.}
\label{mass_rad_figure}
\end{figure*}

Figure \ref{mass_rad_figure} plots mass versus radius for published low-mass stars in EBs with our measurements in red, blue, green, and black circles. For KIC 11922782 and KIC 9821078, our measurements are consistent with the literature. Although the age of KIC 9821078 is not known, when compared to both Dartmouth \citep[][]{Dotter2008} and PARSEC \citep[][]{PARSEC} models with different ages and metallicities, we believe the slight offset of the secondary component is not significant but the primary component is slightly inflated. KIC 11922782 is an old EB system, as was mentioned by \citet[][]{Helminiak2017}. Given the mass of the primary component, it has evolved off of the main-sequence.  We also report the newly measured {\it Kepler} EB, KIC 7605600, whose masses and radii are $M_1 = 0.53 \pm 0.02 M_{\sun}$ and $R_1 = 0.501^{+0.001}_{-0.002} R_{\sun}$ for the primary and $M_2 = 0.17 \pm 0.01 M_{\sun}$ and $R_2 = 0.199^{+0.001}_{-0.002} R_{\sun}$ for the secondary. Both components are low-mass stars and the secondary component is a fully-convective M dwarf star.  The secondary component is one of only a handful fully-convective low-mass stars with empirically measured masses and radii. Combined with KIC 10935310 from \citet[][]{Han2017}, we find all our mass and radius measurements for low-mass {\it Kepler} eclipsing binary stars are consistent with modern stellar evolutionary models for M dwarf stars and do not require inhibited convection by magnetic fields to account for the stellar radii.

With only a handful of {\it Kepler} eclipsing binary stars fully characterized with SB2 radial velocity measurements, it is difficult to draw an overarching conclusion on the nature of radius inflation for all M dwarf stars from these results. However, we can say that we are not seeing the same degree of inflation and scatter in the mass-radius diagram as seen for other eclipsing binary stars, most of which have been analyzed using ground-based photometry and visible-wavelength spectroscopy, whereas our results were obtained using space-based photometry and infrared spectroscopy.  Our results hint at the role of data quality and analysis when reporting eclipsing binary parameters. In any case, \citet{Kesseli2018} presented evidence that fully-convective,
rapidly rotating single M dwarf stars with mass range of 0.08 $M_{\sun}$ $<M<$ 0.18 $M_{\sun}$ are indeed 15-to-20\% larger than evolutionary models predict and stars with mass range of 0.18 $M_{\sun}$  $<M<$ 0.4 $M_{\sun}$ are larger by 6\%, on average. To fully disentangle the nature of magnetic inflation and stellar mass, more low-mass eclipsing binary stars with high signal-to-noise photometry and infrared spectroscopy are required, as well as infrared eclipse photometry to measure individual stars' absolute infrared magnitudes.

\clearpage
\section*{Acknowledgments}
The authors acknowledge support from the NASA Exoplanet Research Program (XRP) under Grant No. NNX15AG08G issued through the Science Mission Directorate. This research involved use of the Discovery Channel Telescope at Lowell Observatory, supported by Discovery Communications, Inc., Boston University, the University of Maryland, the University of Toledo and Northern Arizona University.  This research involved use of the Immersion Grating Infrared Spectrometer (IGRINS) that was developed under a collaboration between the University of Texas at Austin and the Korea Astronomy and Space Science Institute (KASI) with the financial support of the US National Science Foundation under grant AST-1229522, of the University of Texas at Austin, and of the Korean GMT Project of KASI. E.H acknowledges the use of Korean IGRINS time through Korea Astronomy and Space Science Institute (KASI) and thanks Heeyoung Oh for her help with IGRINS observations. E.H thanks Aurora Kesseli for her help with IGRINS observations.

Some of the data presented in this paper were obtained from the Mikulski Archive for Space Telescopes (MAST). STScI is operated by the Association of Universities for Research in Astronomy, Inc., under NASA contract NAS5-26555. Support for MAST for non-HST data is provided by the NASA Office of Space Science via grant NNX09AF08G and by other grants and contracts.  This paper includes data collected by the {\it Kepler} Mission. Funding for the {\it Kepler} Mission is provided by the NASA Science Mission Directorate.  This research involved use of the Massachusetts Green High Performance Computing Center in Holyoke, MA. 

Some of the data presented herein were obtained at the W. M. Keck Observatory, which is operated as a scientific partnership among the California Institute of Technology, the University of California and the National Aeronautics and Space Administration. The Observatory was made possible by the generous financial support of the W. M. Keck Foundation.  This research has made use of the Keck Observatory Archive (KOA), which is operated by the W. M. Keck Observatory and the NASA Exoplanet Science Institute (NExScI), under contract with the National Aeronautics and Space Administration.  The authors wish to recognize and acknowledge the very significant cultural role and reverence that the summit of Maunakea has always had within the indigenous Hawaiian community.  We are most fortunate to have the opportunity to conduct observations from this mountain.

\appendix
\section{Analytic Radial Velocity fitter} \label{RVanalytic}
Fitting radial velocity data requires solving the following equation:
\begin{equation}
    V_P(t) = K_P[\cos{(\omega + \nu(t))} + e\cos{\omega}] + \gamma
    \label{rvequation}
\end{equation}

\begin{equation}
    V_S(t) = - K_S[\cos{(\omega + \nu(t))} + e\cos{\omega}] + \gamma
    \label{rvequation}
\end{equation}

\noindent where $V_P(t)$ and $V_S(t)$ are the radial velocities of the primary and secondary components at time $t$, $K_P$ and $K_S$ are the radial velocity semi-amplitudes of the primary and secondary components, $\omega$ is the argument of periastron, $\nu$ is the true anomaly of the primary component at time $t$, $e$ is eccentricity of the orbit, and $\gamma$ is the systematic radial velocity. Here we use the subscript $P$ and $S$ instead of $1$ and $2$ to denote primary and secondary, and use subscript numbers to indicate radial velocity epochs.  By using $K_P$ and $K_S$, we avoid needing to include the inclination of the orbit.  Equation \ref{rvequation} is a linear function of $K_P$, $K_S$, and $\gamma$ if $\cos{(\omega + \nu)}$ and $e\cos{\omega}$ are known. \\
\indent The high signal-to-noise {\it Kepler} light curves allow us to determine P, $T_0$, $e\cos{\omega}$, and $e\sin{\omega}$ with high precision, much higher than from the radial velocity data alone.  With these parameters from the light curve exclusively, $\nu$ can be determined for any time $t$ by solving Kepler's equation of motion using Newton's method.  Therefore, with the high precision {\it Kepler} data, we linearized the radial velocity model as a function of $K_P$, $K_S$, and $\gamma$, and determined the individual masses analytically.

An analytical solution to the maximum likelihood values of the parameters of interest are calculated using the following matrix multiplication equations:

\begin{align}
    \hat{A} &= \Psi^{-1}G^{T}E^{-1}D \\
    \Psi &= G^{T}E^{-1}G 
\end{align}

\noindent where $\hat{A}$ is the vector of most likely parameters (containing the most likely values of $K_P$, $K_S$ and $\gamma$), $\Psi$ is the parameter covariance matrix, $D$ and $E$ are the data and their covariance matrix, respectively, and G is the basis matrix. The analytic fitter is quick and exact but requires a linear model. \\
\indent Based on the linearization of the radial velocity equation, the basis matrix, G, the data matrix D, and the data covariance matrix E can be formed as shown below to solve for $\hat{A}$ and $\Psi$ analytically.  We assume no covariance between the radial velocity measurements, resulting in $E$ being diagonal.

\begin{gather}
G = 
\begin{bmatrix}
\cos{(\omega + \nu(t_1))} + e\cos{\omega} &  0 & 1\\ 
 \vdots & \vdots & 1\\ 
\cos{(\omega + \nu(t_N))} + e\cos{\omega} &  0 & 1\\ 
 0 & -1 [ \cos{(\omega + \nu(t_1))} + e\cos{\omega}]  & 1\\ 
 \vdots & \vdots & 1\\ 
0 & -1 [ \cos{(\omega + \nu(t_N))} + e\cos{\omega}]  & 1\\ 
 \end{bmatrix}
, D = 
\begin{bmatrix}
V_P(t_1) \\
\vdots \\
V_P(t_N) \\
V_S(t_1) \\
\vdots \\
V_S(t_N) \\
\end{bmatrix}
, E = 
\begin{bmatrix}
\sigma^2_P(t_1) \\
\vdots \\
\sigma^2_P(t_N) \\
\sigma^2_S(t_1) \\
\vdots \\
\sigma^2_S(t_N) \\
\end{bmatrix} \times I \\
\end{gather}

where $N$ represents the number of radial-velocity-measurement epochs, $\sigma_P(t)$ and $\sigma_S(t)$ are the uncertainties on the radial velocity measurements of the primary and secondary at time $t$, and $I$ is an $N$x$N$ identity matrix.  Equations A3 and A4 return the maximum likelihood values of $K_P$, $K_S$, and $\gamma$ and the parameter covariance matrix, $\Psi$, given $G$, $D$, and $E$. \\

\vspace{5mm}
\facilities{DCT (IGRINS), Keck:II (NIRSPEC), IRTF (iSHELL), Kepler}

\software{
\texttt{eb} \citep{Irwin2011},  
\texttt{emcee} \citep[][]{Foreman-Mackey2013},
\texttt{george} \citep[][]{george},
\texttt{mpfit} \citep{Markwardt2009},
\texttt{xtellcor} \citep[][]{xtellcor},
\texttt{REDSPEC} \citep[][]{REDSPEC}
}

\bibliography{references}
\bibliographystyle{aasjournal}

\end{document}